\documentclass[reqno,a4paper,10pt]{amsart}
\usepackage{amsmath,amssymb,amsthm,amscd}
\usepackage[english]{babel}
\usepackage[mathscr]{eucal}

\font\sc=rsfs10 at 12pt

\numberwithin{equation}{section}

\renewcommand{\a}{\alpha}
\renewcommand{\b}{\beta}
\newcommand{\g}{\gamma}

\renewcommand{\d}{\delta}
\newcommand{\D}{\Delta}
\newcommand{\e}{\epsilon}

\renewcommand{\l}{\lambda}
\renewcommand{\L}{\Lambda}
\newcommand{\m}{\mu}
\newcommand{\n}{\nu}

\renewcommand{\r}{\rho}

\renewcommand{\t}{\tau}
\newcommand{\f}{\phi}
\newcommand{\ff}{\varphi}
\newcommand{\F}{\Phi}
\newcommand{\h}{\chi}
\newcommand{\p}{\psi}
\renewcommand{\P}{\Psi}

\newcommand{\C}{{\mathbb C}}
\newcommand{\R}{{\mathbb R}}

\newcommand{\Ab}{{\mathbf A}}

\newcommand{\Bb}{{\mathbf B}}

\newcommand{\Db}{{\mathbf D}}

\newcommand{\Qb}{{\mathbf Q}}

\newcommand{\dF}{\mathfrak d}

\newcommand{\pF}{\mathfrak p}

\newcommand{\Ac}{{\mathcal A}}

\newcommand{\Dcc}{{\mathcal D}}

\newcommand{\Nc}{{\mathcal N}}

\newcommand{\Pc}{{\mathcal P}}

\newcommand{\Dc}{\sc\mbox{D}\hspace{1.0pt}}
\newcommand{\Lcc}{\sc\mbox{L}\hspace{1.0pt}}

\newcommand{\curl}{{\rm curl}\,}

\newcommand{\sgrad}{{\rm sgrad}\,}
\newcommand{\disc}{{\rm disc}\,}
\newcommand{\cont}{{\rm cont}\,}
\renewcommand{\div}{{\rm div}\,}

\newcommand{\supp}{\hbox{{\rm supp}}\,}

\newcommand{\Ker}{\hbox{{\rm Ker}}\,}

\newcommand{\loc}{\operatorname{loc}}

\newtheorem{theorem}{Theorem}[section]
\newtheorem{proposition}[theorem]{Proposition}
\newtheorem{lemma}[theorem]{Lemma}
\newtheorem{corollary}[theorem]{Corollary}

\newtheorem{condition}[theorem]{Condition}
\theoremstyle{definition}

\theoremstyle{remark}

\newtheorem{example}[theorem]{Example}

\begin{document}

\title[Zero modes of the Pauli
operator]{Infiniteness of zero modes for the Pauli
operator with singular magnetic field}
\author[Rozenblum]{Grigori Rozenblum}
\address[G. Rozenblum]{Department of Mathematics \\
                        Chalmers University of Technology \\
                        and University of Gothenburg \\
                        Eklandagatan 86, S-412 96 Gothenburg \\
                        Sweden}
\email{grigori@math.chalmers.se}
\author[Shirokov]{Nikolai Shirokov}
\address[N. Shirokov]{Department of Mathematical Analysis
\\Faculty of Mathematics and Mechanics\\
                      St. Petersburg State University\\2, Bibliotechnaya pl.,
                       Petrodvorets\\
St.Petersburg, 198504, Russia}
\email{matan@math.spbu.ru}

\begin{abstract}
We establish that the Pauli operator describing a
spin-$1/2$ two-dimensional quantum system with a
singular magnetic field  has, under certain
conditions, an infinite-dimensional space of zero
modes, possibly, both spin-up and spin-down, moreover
there is a spectral gap separating the zero eigenvalue
from the rest of the spectrum. In particular,
infiniteness takes place if the field has  infinite
flux, which settles this previously unknown case of
Aharonov-Casher theorem.
\end{abstract}

\date{}

\maketitle
 \tableofcontents

\section{Introduction}
\label{intro}

The presence of zero modes, eigenfunctions with zero
eigenvalues, is a typical feature for two-dimensional
spin $1/2$ quantum systems involving magnetic fields.
Such eigenvalues were first found by Landau (see
~\cite{Landau}) for the Pauli operator with constant
magnetic field, and the multiplicity turned out to be
infinite. Later Aharonov and Casher \cite{AharCasher}
calculated the number of zero modes for a bounded
compactly supported magnetic field, and this number
turned out to be finite and determined by the total
flux of the field. The conditions on the magnetic
field were gradually relaxed, see \cite{Miller,CFKS},
until in \cite{ErdVug} the case of measure-valued
magnetic fields was settled and an Aharonov-Casher
type formula was established for a magnetic field
being a regular measure with finite total variation,
thus producing a finite number of zero modes. On the
other hand, a weak perturbation of the constant
magnetic field leaves the space of zero modes
infinite-dimensional (\cite{Iwatsuka}). In the paper
by Shigekawa \cite{Shigekawa} it was established that
if the field is sufficiently locally  regular and
separated from zero at infinity (or tends at infinity
to zero sufficiently slowly) then, again, the space of
zero modes is infinite-dimensional. On the other hand,
an example in \cite{ErdVug} had shown that if the
total variation of the field is not finite, there may
be no zero modes at all, even if the total flux of the
field, defined as a conditionally convergent integral,
is nonzero. Under some rather restrictive conditions,
infiniteness of zero modes was established for a
periodic magnetic field, see \cite{DubNov80},
\cite{DubNov80a}.

A new type of magnetic fields was recently considered
in relation to the study of zero modes by Geyler and
Grishanov in \cite{GeyGri}. They have studied a system
of  equal Aharonov-Bohm magnetic solenoids placed at
the points of an infinite double-periodical lattice in
the plane. Neither of the previous results apply for
this case, the field being very singular and the total
flux being infinite. Nevertheless, the authors of
\cite{GeyGri} proved that such field produces an
infinite-dimensional zero energy  subspace.  Moreover,
both spin-up and spin-down null subspaces are
infinite-dimensional. This property is proved to be
stable when one adds a constant (positive) magnetic
field, of arbitrary size for the spin-down component,
and not too large for the spin-up component. Further
on, in \cite{GeySto} this result was extended to
certain perturbations of this periodic structure.

Not so much is known about the rest of the spectrum of
the Pauli operator. For the Landau operator, with
constant magnetic field, the spectrum consists of
Landau levels, eigenvalues with infinite multiplicity
placed at the points of an arithmetical progression.
Under a weak perturbation of the field, these
eigenvalues, except the lowest one, may split,
producing a cluster of the discrete spectrum around
the Landau levels (see \cite{Shigekawa}). Thus zero
remain to be an isolated point of the spectrum. On the
other hand, a weak magnetic field without a background
constant field leaves the whole positive semi-axis
belonging to the spectrum, so no spectral gap arises.
If the magnetic field grows unboundedly  at infinity,
the whole spectrum, except   zero, is discrete (see,
again \cite{Shigekawa}). Under rather restricting
conditions the presence of the spectral gap  was
established in \cite{DubNov80}, \cite{DubNov80a} for a
periodic field. However in a more or less general case
this question is still open.

In the present paper we study the zero modes of the
Pauli operator with a non-regular magnetic field with
an infinite total flux. The typical example of the
fields in question is  a, probably infinite, discrete
configuration of AB solenoids on the background of a
more regular magnetic field. The Pauli operator for a
strongly singular field is not essentially
self-adjoint, there is an ongoing discussion on which
self-adjoint extension of the Pauli operator in the
presence of AB solenoids more adequately describes the
real physical situation -- see \cite{AdamiTeta},
\cite{BorgPule}, \cite{Tamura} and  references
therein. It turns out that depending on which
approximation to AB field by more regular fields is
chosen, with simultaneous adjustment of some other
physical parameters, different self-adjoint extensions
can arise. Our main analysis deals with the so called
maximal extension. Its advantage is its  invariance
with respect to singular gauge transformations
reducing the AB fluxes. We handle also another
extension  considered in the paper \cite{ErdVug}, also
gauge invariant, but with different spectral
properties. We discuss the relations of these two
extensions in Sect. 2, as well as describe the
connections of the study of zero modes with problems
in the theory of analytical functions.

  In Sect. 3 we find rather
general conditions for the infiniteness of zero modes
and for zero being an isolated point in the spectrum,
with a possibility to estimate the size of the
spectral gap. We start by settling  the long-standing
hypothesis (see the discussion in \cite{ErdVug}) that
a field of constant direction with infinite total flux
produces infinitely many zero modes. Further on, we
 show that this infiniteness is preserved under addition of a field with
 different direction, having a finite flux. This
 establishes Aharonov-Casher theorem for the case of an
 infinite total flux of the field.
 We pass then to the case when this 'wrong' component
 may have an infinite flux. Here, we suppose that
the flux of the field through any disk of a fixed size
is non-negative, at least far enough from the origin,
 moreover the flux of the {\em averaged} field is
infinite. This requirement, together with some
additional local conditions, grants that the spin-down
zero subspace is infinite-dimensional.  If,
additionally, the above local fluxes are separated
from zero, then zero is an isolated point of the
spectrum of the Pauli operator. For
 regular fields, this condition  prevents the infiniteness of the
 spin-up
 zero modes, since spin inversion corresponds to changing the sign of
  the field. However, if the discrete component of the field is large
   enough, in other words, if sufficiently many Aharonov-Bohm solenoids are present,
   then it turns out that the main condition can be
   satisfied for the spin-up component of the 'maximal' operator as well, so
there are infinitely many spin-up zero modes too. We
also explain how the results change when we pass to
the self-adjoint Pauli operator considered in
\cite{ErdVug}.  Here the situation with both spin-up
and spin-down zero modes does not appear. We conclude
Sect.3, by some examples, in particular the case of a
periodic and quasi-periodic magnetic field fits into
the general approach, and the results of
\cite{DubNov80} and \cite{DubNov80a} are substantially
extended.

Further on, we pass to the situation when the general
results are not sufficient, since the main condition
of positivity of local flux may be violated. Supposing
that the magnetic field in question is a perturbation
of some initial field where a quadratic lower estimate
for the potential is known, we establish such
quadratic estimate for the potential of the perturbed
field, thus ensuring the infiniteness of zero modes,
but, probably, without the spectral gap. Among others,
 the constant one,
AB-lattice, probably, on the background of a constant
field, a periodic or quasi-periodic field with some
mild local regularity may serve as the unperturbed
field. Admissible perturbations are rather general, in
particular they allow existence  of arbitrary large
regions on the plane with field having 'wrong'
direction. In the end we discuss some examples where
the perturbation theorems can be applied.  The
estimates for the potential obtained on this way of
reasoning, may be useful in the further study of the
perturbation of the Pauli operator by an electric
field.

The starting point of our study was an attempt to
understand the possibility of perturbing  the results
of \cite{GeyGri}, by means of changing the intensities
and positions of the AB solenoids. We thank V. Geyler
who attracted our attention to this kind of problems.
 Further on, when it turned out that much more general situations can be
 taken care of,  the proof of the crucial theorem 3.2 appeared in the process of
discussions with  F. Nazarov. We highly appreciate
also the discussions with B. Berndtsson on the
spectral gaps and with L. Erd\"os about the definition
of the Pauli operator. The second author (N.Sh.) was
supported by the stipend from the Swedish Royal
Academy of Sciences.  Both authors thank the
Mittag-Leffler Institute for hospitality when the work
on the paper was in its most active phase.

\section{Definition of the operator}
\label{Defin}

 We  identify the
real two-dimensional space $\R^2$ with co-ordinates
$x=(x_1,x_2)$ with the complex plane $\C$, setting
$z=x_1+ix_2$; as usual,
$\bar{\partial}=\partial_{\bar{z}}=(\partial_1+i\partial_2)/2$,
${\partial}=\partial_{{z}}=(\partial_1-i\partial_2)/2,
\partial_j=\partial_{x_j}$, and the Lebesgue measure
will be denoted by $dx$.

Formally, the Pauli operator in $L_2(\R^2)$, with
gyro-magnetic ratio $g=2$, is defined as the square of
the Dirac operator \begin{footnote}{Different sign
conventions are used in the literature. We follow the
sign choice made in \cite{ErdVug}, which is the
opposite to the one made in, say,
\cite{CFKS}.}\end{footnote}
$$
    \Dcc=\sigma\cdot(-i\nabla+\Ab)=(\sigma_1(-i\partial_1+A_1)+\sigma_2(-i\partial_2+A_2)).
$$
Here $\sigma_1, \sigma_2$ are the Pauli matrices,
$
\sigma_1=\left(%
\begin{array}{cc}
  0 & 1 \\ 1 & 0 \\
\end{array}%
\right), \:\: \sigma_2=\left(%
\begin{array}{cc}
  0 & -i \\ i & 0 \\
\end{array}%
\right),$ and $A_j$ are real functions, components of
the {\it magnetic potential} $\Ab=(A_1,A_2)$. So,
$
    \Pc=\Dcc^2=-(\sigma\cdot(\nabla+i\Ac))^2.
$

Introducing the notations $\Pi_j=-i\partial_j+A_j$,
$Q_{\pm}=\Pi_1\pm i \Pi_2$,  we can represent the
Pauli operator $\Pc$ as
\begin{equation}\label{2:Paulimatrixform}
   \left(\begin{array}{cc}
     \Pc_+& 0 \\
     0 & \Pc_-  \\
   \end{array}\right)= \left(\begin{array}{cc}
     Q_-Q_+ & 0 \\
      0 & Q_+Q_- \\
    \end{array}
\right).
\end{equation}
Formally, the operators $Q_{\pm}$ are adjoint to each
other.

The {\it magnetic field} is defined as
$
    \Bb=\curl \Ab=\partial_1A_2-\partial_2A_1,
$ and it  is considered in the classical physics as
the only actual physical reality, the potential being
merely a mathematical fiction. This is the fact also
in the quantum physics, provided the magnetic field
(and therefore the potential) are not too singular.
The latter statement means that  if for two magnetic
potentials $\Ab_1, \Ab_2$  the equality
$\Bb=\curl\Ab_1=\curl\Ab_2$ holds in the proper
distributional sense then the corresponding Pauli
operators are gauge equivalent: there exists a real
function $\phi$ such that the multiplication by
$\exp(i \phi)$ transforms one of the corresponding
Pauli operators into another. To assign a rigorous
meaning to the above statement, one has to define the
Pauli operator as a self-adjoint operator in the
Hilbert space $L_2(\R^2)$, with a certain domain. The
gauge transformation, the multiplication by $\exp(i
\phi)$, should transform differential expression in
the proper way as well as transform the domain of one
operator to the domain of the other one.

The standard definition of the Pauli operator by means
of quadratic forms requires $\Ab\in L_{2,\loc}$ and is
described, for example in \cite{ErdVug} or \cite{Sob}.
Having in mind the  representation
\eqref{2:Paulimatrixform} of the Pauli operator, the
quadratic form
\begin{equation}\label{2:Pauliqform}
    \pF_\Ab[\p]=\int|\sigma\cdot(-i\nabla+\Ab)\p|^2dx=\|Q_+\p_+\|^2+\|Q_-\p_-\|^2,
    \p=\left(\begin{array}{c}
      \p_+ \\
      \p_-\\
    \end{array}\right)
\end{equation}
is introduced. If the magnetic potential $\Ab$ is
sufficiently regular, say, $\Ab\in L_{4,\loc}, $ $
\curl\Ab\in L_{2,\loc}, \div\Ab\in L_{2,\loc}$, the
quadratic form \eqref{2:Pauliqform} can, actually, be
obtained from the expression $(\Pc \p,\p)$, $\p\in
C_0^\infty$ by means  of  the integration by parts,
thus justifying the use of  \eqref{2:Pauliqform} for
more singular potentials. So, for $\Ab\in L_{2,\loc}$
one can chose some domain $\dF$ for the form
$\pF_\Ab$, where this form is closed, and accept the
self-adjoint operator corresponding to this form as
the Pauli operator. Unlike the case of the magnetic
Schr\"odinger operator, where all reasonable choices
of the domain of the form turn out to be equivalent
(see \cite{Simon}), for the Pauli operator such
equivalence is not established.  In \cite{ErdVug} the
authors argue that the choice of the {\em maximal}
domain consisting of all functions $\p\in L_2$ for
which \eqref{2:Pauliqform} is finite, is physically
reasonable since this corresponds to the states with
finite energy.

 For
magnetic fields possessing local singularities, the
ones  which we are going to study further on, this
description is not satisfactory, as it was explained
in \cite{ErdVug}, and another approach, based upon the
scalar potential, was proposed. We will use the same
way of defining the operator, with certain
modifications. In what follows, the magnetic field
will be represented by a Borel {\em signed} measure
$\mu$ having locally finite variation. We suppose,
moreover, that the support of the discrete part of the
measure does not have finite accumulation  points. To
the field $\mu$ we associate a {\em scalar potential}
$\Psi(x)$, a solution of the equation $
    \Delta\Psi=\mu
$
in the sense of distributions. The corresponding
vector potential is defined as
$
    \Ab=(A_1,A_2)=\sgrad\Psi=(\partial_2\Psi,-\partial_1\Psi),
$
again in the sense of distributions.

The quadratic form \eqref{2:Pauliqform}, under certain
regularity conditions, can be transformed to
\begin{equation}\label{2:PauliErdosForm}
    \pF[\psi]=\pF_+[\psi_+]+\pF_-[\Psi_-]=4\int|\bar{\partial}(e^{-\Psi}\psi_+)|^2e^{2\Psi}dx
    +4\int|{\partial}(e^\Psi\psi_-)|^2e^{-2\Psi}dx.
\end{equation}
For a field $\mu$ with singularities, is is the form
\eqref{2:PauliErdosForm} that is used for defining the
Pauli operator. The decomposition of the measure
$\mu$, $\mu=\mu_\cont+\mu_\disc$ leads to a similar
decomposition of the potential,
$\Psi=\Psi_\disc+\Psi_\cont.$ We will use  the
potential $\Psi_\cont$  constructed in \cite{ErdVug}.
 This is a function satisfying the equation
    $\D\Psi_\cont=\mu_\cont$
 in the sense of distributions. It is established in
\cite{ErdVug} that such a function exists and
possesses certain regularity properties, in
particular,
 $
    \exp(\pm2\Psi_\cont)\in L_{1,\loc}(\R^2),\;
    \nabla\Psi_\cont\in L_{p,\loc},
    p<2.$

If only the continuous part of the measure is present,
the natural domain for the form
\eqref{2:PauliErdosForm} consists of all functions
$\psi_\pm$ for which  \eqref{2:PauliErdosForm} is
finite. Although this definition is rather implicit,
this domain possesses an easily describable core: the
space of functions $\psi_\pm$ for which
$e^{-\Psi}\psi_+$ and $e^{\Psi}\psi_-$ are smooth
functions with compact support (see \cite{ErdVug}).

We assume next that only the discrete part of the
measure $\mu=\mu_\disc$ is present,
\begin{equation}\label{2:DiscretePart}
    \mu_\disc=2\pi\sum_{\l\in\L}\a_\l \d(x-\l),\;
    x\in\R^2.
\end{equation}
The support $\L$ of $\mu_\disc$ will be supposed to be
a discrete set, without finite accumulation points,
moreover, {\em uniformly discrete}:
\begin{equation}\label{2:sparse}
    \mbox{dist}(\l, \L\setminus\l)\ge r^0,
    r^0>0,\mbox{ for any } \l\in\L.
\end{equation}
Each component of the discrete measure is an
Aharonov-Bohm (AB) solenoid (see \cite{AharBohm}) with
flux $2\pi\a_\l$ and intensity $\a_\l$. We consider
the case of one solenoid first. The AB magnetic
potential  corresponding to one term
 in \eqref{2:DiscretePart},
$\Bb=2\pi\a\d(x-x^0);\; x^0=(x^0_1,x^0_2)$,
with intensity $\a$ can be chosen as
  $  \Ab(x)=\left(-\a\frac{x_2-x_2^0}{r^2}, \a\frac{x_1-x_1^0}{r^2}\right),
    r=|x-x^0|.$
 The corresponding scalar potential is
   $ \Psi(x)=\Psi(z)={\a}\ln|z-z^0|\;$
(here $z^0=x_1^0+x_2^0,$ from now on, the complex
picture is more convenient.) Thus the expressions
$\exp(\pm\Psi)$ have singularities of the form
$|z-z^0|^{\pm\a}$ at $z^0$.

Formally, the Pauli operator with AB field admits
gauge transformations. For an integer $m$, we set
$\f(z)=\exp(-im \arg(z-z^0))$. Then the multiplication
by $\f$ transforms the AB Pauli operator with
intensity $\a$ to the one with intensity $\a+m$. The
potential correspondingly transforms as $\Psi\mapsto
\Psi|z-z^0|^m$. Whether this transformation is a
unitary equivalence of operators depends on how the
self-adjoint operator corresponding to the form
\eqref{2:PauliErdosForm} is defined.

To make our description of such self-adjoint operators
more precise, we introduce the following notations. In
what follows, the notations $\bar{\partial}, \partial$
have the meaning of derivatives in the sense of the
space of distributions $\Dc'(\R^2)$. For a closed set
$E$, we denote by $\bar{\partial}_E, \partial_E$ the
derivatives in the sense of $\Dc'(\R^2\setminus E)$.

We explain now the way of defining the Pauli operator,
proposed in \cite{ErdVug}. For $\a\in[-1/2,1/2)$, one
accepts as a domain of the form
\eqref{2:PauliErdosForm} the space of such functions
$\psi\in L_2(\R^2)$ for which the derivatives
$\bar{\partial}(e^{-\Psi}\psi_+)$ and
${\partial}(e^{\Psi}\psi_-)$ (thus taken in the sense
of distributions in $\Dc'(\R^2)$) are functions, and
the form is finite:
\begin{equation}\label{2:DomEV}
    \pF[\psi]<\infty.
\end{equation}
 With such domain, which we denote
here by $\dF_{EV}(\a)$, the form
\eqref{2:PauliErdosForm}  is closed and defines the
self-adjoint operator which we denote by
$\Pc_{EV}=\Pc_{EV}(\a)$. For $\a$ outside the above
interval, the operator is defined by means of the
gauge transformation. For a given $\a$, let $\a^*$ be
the unique number in the interval $[-1/2,1/2)$ such
that $\a^*-\a=m$ is an integer. Then the Pauli
operator   $\Pc_{EV}(\a)$ is defined as
  $\Pc_{EV}(\a)=
  \exp(im\arg(z-z^0))\Pc_{EV}(\a^*)\exp(-im\arg(z-z^0)).$
With this definition, the operator is automatically
gauge invariant. However, for $\a\notin[-1/2,1/2)$,
the description of the domain does not agree with
\eqref{2:DomEV}. In fact, if the distributional
$\bar{\partial}(e^{-\Psi}\psi_+)$ is a function, for
$\Psi=|z-z^0|^{\a^*}$, the gauge transformation leads
to the expression
$\bar{\partial}((z-z^0)^me^{-\Psi}\psi_+)$, which is
not necessary  a function, it may contain the
$\delta$-distributions and its derivatives. This might
be considered as a minor inconvenience, however it
leads to the unnatural absence of invariance of the
number of zero modes under the change of sign of the
magnetic field, as can be seen from the version of the
Aharonov-Casher theorem in \cite{ErdVug} (or, more
easily, from the non-symmetry of the main interval
$[-1/2,1/2)$, chosen arbitrarily  - see \cite{MicPer}
for more details).

We consider, along with the above operator, an
alternative one. For a given $\a$ we define
$\Pc_{\max}=\Pc_{\max}(\a)$ as the operator
corresponding to the quadratic form
\begin{eqnarray}\label{2:Pauliformrestricted}
\!\!\!\!\!\pF_{\max}[\psi]=\pF_{+\max}[\psi_+]+\pF_{-\max}[\psi_-]\;\;\;\;\nonumber\\
=4\int|\bar{\partial}_{\{z^0\}}(e^{-\Psi}\psi_+)|^2e^{2\Psi}dx
    +4\int|{\partial}_{\{z^0\}}(e^\Psi\psi_-)|^2e^{-2\Psi}dx.
\end{eqnarray}
defined on such functions $\psi$ that the derivatives
in \eqref{2:Pauliformrestricted} (thus understood in
the sense of $\Dc'(\R^2\setminus \{z^0\})$) are
functions and $\pF_{\max}[\psi]$ is finite. The
operator $\Pc_{\max}$ is again gauge invariant (see,
again, \cite{MicPer} for corresponding calculations).

Both constructions can be carried over to the case of
a finite or infinite system of AB solenoids placed at
the points of  a discrete set $\L$ of the plane, as in
\eqref{2:DiscretePart}. We say that the
vector-function $\Ab(x)=(A_1(x),A_2(x))$ is a vector
potential for the magnetic field
\eqref{2:DiscretePart} if
   $ \mu_\disc=\curl \Ab$
in the sense of $\Dc'(\R^2)$. The function
$\Psi_{\disc}$ satisfying the Poisson equation
$\D\Psi_{\disc}=\mu_{\disc}$ is the scalar potential.
As above,  we define the  quadratic form $\pF$ by the
expression in \eqref{2:PauliErdosForm}. The gauge
transformations enable changing all intensities
$\a_\l$ by arbitrary integers. If $m_\l, \l\in\L,$ is
a collection of integers then the gauge transformation
changing the intensity at the point $\l$ by $2\pi
m_\l$ consists in the multiplication by a function
$W(z)$. The function $W(z)$, as proposed in
\cite{ErdVug}, equals $\frac{L(z)}{|L(z)|}$,
$L(z)=F(z)\bar{G}(z)$ where $ F(z)$ is an analytical
function having zeros of order $m_\l$ at the points
$\l$ where $m_\l>0$, and $G(z)$ has zeros of order
$-m_\l$ at the points $\l$ where $m_\l<0$.
 For any collection of $\a_\l$, by adding proper
 $m_\l$, one obtains  the {\em reduced} intensities
 $\a^*_\l\in[-1/2,1/2)$, used for defining the
 operator. As in the case of a single AB solenoid,
 the operator $\Pc_{EV}(\{\a_\l,\l\in\L\})$ is defined as the one gauge
 equivalent to $\Pc_{EV}(\{\a_\l^*,\l\in\L\})$, the latter
 determined by the quadratic form
 \eqref{2:PauliErdosForm} on all functions $\psi$ for
 which this form is finite. Alternatively, the {\em
 maximal} operator $\Pc_{\max}(\{\a_\l,\l\in\L\})$ is defined for any set of (non-integer) $\a_\l$
 by the quadratic form $\pF_{\L}$ in \eqref{2:Pauliformrestricted}
 on the functions for which this latter form is
 finite.

 Finally, when both discrete and continuous components
 of the measure $\mu$ are present, the discrete one
 located at the points $\l$ of a discrete set $\L$,
 the Pauli operators $\Pc_{EV}$ and $\Pc_{\max}$ are
 defined in a similar way, with only difference that
 the scalar potential $\Psi$ is now the sum of the
 potentials $\Psi_\disc$ and $\Psi_\cont$
 corresponding to the discrete and continuous parts of
 the measure $\mu$. We do not
 touch upon the question on which of these operators
 (if any) describes the actual physical picture. we
 keep however in mind that for a continuous measure
 as a field these operators coincide.

 In the general situation it is hard to describe the
 domain of these two operators explicitly.
 However, more can be said about the null subspace of
 these operators, in other words, about the zero modes.
Note, first of all, that the quadratic form of
$\pF_{\max}$
 is an extension of the form $\pF_{EV}$. Since both
 forms are non-negative, this implies that
 $\Ker(\Pc_{EV})\subset\Ker(\Pc_{\max})$.  This can
 also be seen from the direct description of the zero
 modes. If a function $\psi=(\psi_+, \psi_-)$ lies in the null
 subspace of the operator $\Pc_{EV}$ or $\Pc_{\max}$
 than $\psi$ must annule the corresponding quadratic
 forms $\pF_{EV}$, $\pF_{\max}$, which means
 \begin{equation}\label{2:CREV}
\bar{\partial}(e^{-\Psi}\psi_+)=0,
{\partial}(e^{\Psi}\psi_-)=0
\end{equation}
for  $\Pc_{EV}$
 and
 \begin{equation}\label{2:CRmax}
\bar{\partial}_\L(e^{-\Psi}\psi_+)=0,
{\partial}_\L(e^{\Psi}\psi_-)=0
\end{equation}
for $\Pc_{\max}$. Both \eqref{2:CREV} and
\eqref{2:CRmax} mean that the function
$f_+=e^{-\Psi}\psi_+$ must be analytical,
$f_-=e^{\Psi}\psi_-$ must be anti-analytical, but on
different sets. For the operator $\Pc_{EV}$, by
\eqref{2:CREV}, these functions must be entire
functions of variables $z,\bar{z}$ respectively. On
the other hand, for the operator $\Pc_{\max}$, by
\eqref{2:CREV} these functions may have poles at the
points $\l\in\L$, but not too strong ones, so that
still after the multiplication by $\exp(\pm\Psi)$,
they get into $L_2$.

To make things more concrete we suppose that from the
very beginning the gauge transformation is made, so
that all intensities $\a_\l$ are  in the interval
$(0,1)$ for $\Pc_{\max}$. In this case,  the function
$e^\Psi$ behaves as $|z-\l|^{\a_\l}$ near the point
$\l\in\L$.  Therefore the condition $\partial_zf_-=0$,
together with $f_-\exp(-\Psi)\in L_2$ leads to
anti-analyticity of $f_-$ at the points of $\L$ as
well: near a point
 $\l\in\L$ the function $f_-^2$ must be summable with weight
   having a singularity of the form
 $|z-\l|^{-2\a_\l}$, therefore the  possible singularity of $f_-$
 is removable. On the other hand, for the spin-up component $\psi_+$, the function
$f_+$
 has to be holomorphic outside $\L$ but near the
 points of $\L$ it must belong  to $L_2$ with weight
 $|z-\l|^{2\a_l}$, which, due to $\a_\l\in(0,1)$, allows $f_+$ to have a
 simple pole at $\l$. This asymmetry  can be
 reversed by changing the normalization of the
 discrete part of the measure: by mean of a gauge
 transformation we can decrease all intensities by $1$
 thus arriving at the measure $\m'$ with negative
 discrete part having the intensity $\a'_\l=(\a_\l-1)\in(-1,0)$
 at the point $\l\in \L$. Then it is for the spin-up
 component that the null subspace is generated by entire
 functions, and for the spin-down one by meromorphic
 functions with simple poles. We can, moreover,  take
 the first normalization when studying the spin-down component
 and the second one for the spin-up component, thus
 only entire functions will be  involved. With this
 last
 agreement accepted,
  the derivatives involved in the forms
do not depend in the space of distributions where they
are considered, so we can painlessly omit the
corresponding subscripts in our notations.

To compare the null subspaces of two operators under
consideration, we suppose first that all intensities
$\a_\l$ lie in $(0,\frac12)$. In this case for the
spin-down component $\Pc_-$ null subspaces coincide,
being in both cases generated by entire
anti-analytical functions. The  null subspace of
$\P_{+\max}$ may be larger than the  null subspace of
$\P_{+EV}$, since the latter subspace may involve
meromorphic functions $f_+$ in addition to entire
functions for $\Pc_{+EV}$.  If some of the intensities
$\a_\l$ lie in $[-\frac12,0)$, both $\Pc_{+\max}$ and
$\Pc_{-\max}$ may have zero modes generated by
meromorphic functions, so both null subspaces may turn
out to be larger than the ones for $\Pc_{EV}$.  See,
again, \cite{MicPer} for more detailed comparison of
these two self-adjoint extensions.

Now we  make  some more observations about the part
$\Psi_\disc$ of the potential, the one responsible for
the discrete part of the measure. It follows from the
uniform discreteness condition \eqref{2:sparse}
  that the discrete  set $\L$  has a density not higher than
that of a regular lattice, more exactly,
\begin{equation}\label{2:Density}
   N(R)\equiv\sharp\{\l\in\L,\l<R\}=O(R^2),r\to\infty.
\end{equation}
 Consider the sum
\begin{equation}\label{2:Potential}
\Psi_\disc(z)=\a_0\log|z|+
\sum\limits_{\l\in\L,\l\ne0}\a_\l\left(\log|1-\frac{z}\l|+
\Re\left(\frac{z}\l+\frac12\left(\frac{z}\l\right)^2\right)\right);
\end{equation}
if $\l=0$ does not belong to the set $\L$, the first
term in \eqref{2:Potential} is omitted. The series
converges uniformly on any compact set in $\C$ not
containing the points in $\L$, moreover the Laplace
operator can be applied term-wise, so
\eqref{2:Potential} produces the required potential.

 The particular case of a special
interest is the one of a purely discrete measure, a
{\it regular} lattice, with all intensities equal,
\begin{equation}\label{2:RegularLattice}
    \L=\{\l_{m_1m_2}\}=\omega_1 m_1+\omega_2 m_2,\;\a_{\l_{m_1m_2}}=\a\in(0,1),
\end{equation}
where $ \omega_1,\omega_2$ are complex numbers with
non-real $\omega_1/\omega_2$. In for this
configuration, as it was noticed in \cite{GeyGri} the
potential $\Psi$ is closely related to  the
Weierstrass $\sigma$-function
\begin{equation}\label{2:Weierstrass}
    \sigma(z)=z\prod\limits_{\l\ne0,\l\in\L}
    \left(1-\frac{z}{\l}\right)\exp\left(\frac{z}\l+\frac12\left(\frac{z}\l\right)^2\right),
    \end{equation}
so that
$\tilde{\Psi}_0(z)=\a\log|\sigma(z)|$
can serve as a  potential for the magnetic field
\eqref{2:DiscretePart}. It was established in
\cite{Perelomov} that, the potential $\tilde{\Psi}_0$
possesses a very special property:
\begin{equation}\label{2:periodicity}
   \tilde{\Psi}_0(z)=\a(\Re(\n z^2)+m |z|^2 +\r(z)),
\end{equation}
where $\n$ is a certain coefficient determined by the
lattice, $m=\frac\pi{2S}$, $S$ being the area of the
elementary cell, and $\r$ a $\L$-periodic function,
with proper logarithmic singularities at the points of
the lattice. We do not care about the value of $\n$,
explicitly given in \cite{Perelomov,GeyGri}. Anyway,
the first summand in \eqref{2:periodicity} is a
harmonic function; we subtract it and for a regular
lattice we will consider
\begin{equation}\label{2:realperiodicity}
\Psi_{0}(z)=\a(m |z|^2 +\r(z)),
\end{equation}
with  $\r$ determined by \eqref{2:periodicity}. With
such potential $\Psi_{0}$, for any entire function
$f({z})$ subject to $|f({z})|\le C\exp(\g|z|^2)$ with
$\g<\a m$, the function $\psi_-(z)=\exp(-\a(m |z|^2
+\r(z)))\bar{f}({z})$ belongs to $L_2$. This
observation made in \cite{GeyGri} proves infiniteness
of zero modes for  $\Pc_{-\max}$. Passing to the
operator $\Pc_+$, we  make the gauge transformation
reducing all fluxes to $2\pi (\alpha-1)$. For the
reduced operator, the function $\Psi_{0+}(z)=(\a-1)(m
|z|^2 +\r(z))$ serves as a potential, and thus any
entire function  $f_+$ of the variable ${z}$ produces
the $L_2$ zero mode $\psi_+=\exp((\a-1)(m |z|^2
+\r(z)))f_+$ of the operator $\Pc_{+\max}$. As it is
explained above, if  $\a$ lies in the interval
$(0,\frac12)$, infiniteness of zero modes holds also
for $\Pc_{-EV}$, and the operator $\Pc_{+EV}$, as
follows easily from the properties of $\Psi_{0}$, has
no zero modes. In the case $\a\in[-\frac12,0)$ the
spin-up and spin-down components change their roles.
In Sect.4,  we will show that such estimates for the
potential are preserved under certain types  kinds of
the perturbations of the regular AB lattice, thus, in
particular, providing us with examples of  magnetic
fields possessing arbitrarily large regions with
'wrong' direction of the field, but, nevertheless,
with infinitely many zero modes.

So, under our normalization conditions, the study of
zero modes is reduced to the study of existence of
entire functions which, after being multiplied by a
certain weight get into $L_2$. This study can be done
by means of explicit estimates for the potential
$\Psi$, like in
\cite{AharCasher,CFKS,Miller,ErdVug,GeyGri} or by
indirect methods, cf. \cite{Shigekawa}. We are going
to combine both approaches.

\section{Zero
modes and the spectral gap. Methods of the theory of
subharmonic functions}\label{sect:7inf}

In this section we establish the  infiniteness of zero
modes  under rather general conditions. We start by
proving this for the magnetic field with constant sign
and infinite flux, and then relax  the  positivity
restriction in different ways. Further on we find
conditions for the existence of the spectral gap.

We suppose that the general conditions on the measure
as  formulated in the previous section are fulfilled.

\begin{theorem}\label{3:TheoPos}
Let  $\mu$ be non-negative locally finite Borel
measure on $\C$, with the support of the discrete part
not having finite accumulation points, and the
normalization agreements of the previous Section be
fulfilled. Then the operators $\Pc_{-EV}$,
$\Pc_{-\max}$ have infinitely many zero
modes.\end{theorem}

As it is explained in the previous section, it is
sufficient to establish the following fact about
(anti-)analytical functions, which is valid for {\em
any} non-negative measure.  Theorem~\ref{3:TheoPos}
follows from it, with an obvious replacement of $\mu$
by $2\mu$.

\begin{theorem}\label{3:TheoSubhar} Let $\P$ be a
subharmonic function such that for the measure
$\mu=\Delta\P$,
\begin{equation}\label{3:inftness}
\mu(\C)=\int_\C d\mu(z)=\infty.
\end{equation}
Then the spaces of entire analytical and
anti-analytical functions $f$ such that
\begin{equation}\label{3:L2}
    \int_\C |f(z)|^2 e^{-\P}dx<\infty
\end{equation}
are infinite-dimensional. \end{theorem}

\begin{proof} Of course, it  suffices to establish just one part,
 say, about analytical functions, which we are going to do,
 for convenience of references. The strategy of proving the theorem is
the following.  We chose a sequence of points $z_n,\;
n=1,\ldots,$ in a special way.  For any given $n$ a
collection  of functions $f_k,\; k=1,\ldots,n$
satisfying \eqref{3:L2} will be constructed in such
way that $f_k(z_l)=0,\; k<l, f_k(z_k)\ne0$.  Such
system of functions is, obviously, linearly
independent.

We denote by $\Db(z,R)$ the open disc centered at $z$
with radius $R$. All the measures in the proof are
supposed to be non-negative.

We take $z_1=0$. Then we chose $R_1$ so that
$\mu(\Db(0,R_1))\ge 46$. Then for each $k>1$ we find
$R_k$ so that $\mu(\Omega_k)\ge 20,$
$\Omega_k=\Db(0,R_k)\setminus\Db(0,R_{k-1})$,
$\Omega_1=\Db(0,R_1)$. This can be done due to
infiniteness of \eqref{3:inftness}. Then we take any
$z_k$, strictly inside $\Omega_k$,   not in the
support of $\mu_\disc$, $k\ge2$.

Now we fix $n$ and find a sufficiently small positive
$\d$ such that the disks $\Db(z_k,\d)$ $k=1,\ldots,n$
lie strictly inside respective $\Omega_k$. Then we fix
measure $\mu_0\le\mu$ supported in $\Omega_1$ such
that
$\mu_0(\Omega_1)=26.$
 Denote by $\P_0$ the logarithmic
potential of measure $\mu_0$,
    $\P_0(z)=(2\pi)^{-1}\int \ln|z-w|d\mu_0(w).$
 The function $\P_0$ behaves  as
$\P_0(z)\sim\frac{13}{\pi}\ln|z|$ as $|z|\to\infty$.

We set further $\P_1=\P-\P_0$, $\mu_1=\mu-\mu_0\ge0$,
so $\Delta\P_1=\mu_1$ and, by our construction,
\begin{equation}\label{3:Psi1}
    \mu_1(\Omega_k)\ge20, k=1,\ldots,n.
\end{equation}
Further on we chose measures $\n_k\le\mu_1,
k=1,\ldots,n,$ supported in respective $\Omega_k$,
 and such that
$\nu_k(\Omega_k)=20$, and denote by $U_k$ the
logarithmic potential of the measure $\nu_k$, with the
same asymptotic behavior $U_k\sim\frac{10}{\pi}\ln|z|$
for large $|z|$. For a   positive $h<\d/2$, we denote
by $\n'_k, k=1,\ldots,n,$ the measures supported in
the respective disks $\Db(z_k,h)$ and  coinciding
there with $\frac{20}{\pi h^2}$ times the Lebesgue
measures, so that the logarithmic potentials $U_k'$ of
these measures have the same asymptotic behavior for
large $|z|$ as $U_k$,
$U'_k(z)\sim\frac{10}{\pi}\ln|z|$. We denote by $
U^h(z)$ the subharmonic function
\begin{equation}\label{3:tildeU}
{U}^h(z)=\P_1(z)+\sum_{k=1}^n (U_k'(z)-U_k(z)).
\end{equation} By our choice of measures,  $U_k$ and $U_k'$ differ controllably for large $z$.
In fact,
\begin{equation}\label{3:difference}
    U_k'(z)-U_k(z)=\frac{1}{2\pi}\int\limits_{\Omega_k}\ln\left|
    1-\frac{w}{z}\right|(d\n'_k-d\n_k),
\end{equation}
and since for $|z|\ge 2(R_n+\d)$, $|w|<R_n$, we have
$\ln|1-w/z|<\ln2$, so
$|U_k'(z)-U_k(z)|<20\ln2(2\pi)^{-1}<15$. Adding up
such estimates for all $k$, we obtain
\begin{equation}\label{3:Corrected Potential}
    |U^h(z)-\P_1(z)|\le 15n
\end{equation}
for large $|z|$, $|z|\ge 2(R_n+\d)$. Therefore, for
any  non-negative function $v$, and any $R\ge
2(R_n+\d)$,
\begin{equation}\label{3:Weighted estimate1}
    \int_{R\le |z|\le 2R} v\exp(-\P_1)dx\le e^{15n} \int_{R\le |z|\le 2R}
    v\exp({-{U}^h})dx,
\end{equation}
For  $z$ small, $|z|\le 2(R_n+\d)$ but lying outside
the disks $\Db(z_k, \d)$, we note that each function
$U'_k(z)$, being the logarithmic  potential of a
measure supported in the disk $\Db(z_k, h), \;
h\le\d/2$, is bounded by some constant depending on
$\d$  and $R_n$, $|U'_k(z)|\le C(d, R_n)$. The
potential $U_k$, being the logarithmic potential of a
measure supported in the disk $\Db(0,R_n)$, is not
necessarily bounded from below but it is bounded from
above, again, by some constant depending on $\d$  and
$R_n$, $U'_k(z)\le C(\d, R_n)$  for $|z|\le
2(R_n+\d)$.
 This gives us
 \begin{equation}\label{3:CorrectedPotential2}
U^h(z)-\P_1(z)=\sum (U_k'(z)-U_k(z))\ge -2nC(\d,R_n),
\; |z|\le 2(R_n+\d), z\notin\cup\Db(z_k,\d).
\end{equation}

Next we fix a function $\ff\in C_0^\infty(\C)$ such
 that $\ff$ vanishes in the disks $\Db(z_k,\d)$,
 $k<n$,
 but $\ff(z)=1$ in $\Db(z_n,\d).$
  To the function
 $\bar{\partial}\ff$ we apply the theorem by
 H\"ormander,
 see \cite{Hormander}, Theorem 4.4.2, on
 the solutions of $\bar{\partial}$-equation in weighed
 spaces: we find a function $g=g_h$ solving the equation
 $\bar{\partial} g_h=\bar{\partial}\ff$ such that
\begin{equation}\label{3:weighted Estimsolutions1}
    \int_{\C}|g_h|^2\frac{e^{-{U}^h}}{(1+|z|^2)^2}dx\le
\int_{\C}|\bar{\partial}\ff|^2{e^{-{U}^h}}dx.
\end{equation}
We recall now that $\bar{\partial}\ff(z)=0$ in the
disks $\Db(z_k,\d)$. Therefore the estimate
\eqref{3:weighted Estimsolutions1} gives
\begin{equation}\label{3:weighted Estimsolutions2}
\int_{\C}|g_h|^2\frac{e^{-{U}^h}}{(1+|z|^2)^2}dx\le
\int_{\C\setminus\cup\Db(z_k,\d)}|\bar{\partial}\ff|^2{e^{-{U}^h}}dx.
\end{equation}

Now we use the estimates \eqref{3:Corrected Potential}
and \eqref{3:CorrectedPotential2} which enable us to
replace in the right-hand side of \eqref{3:weighted
Estimsolutions2} the weight $\exp(-{U}^h)$ by
$\exp(-\P_1)$. We obtain therefore the inequality
\begin{equation}\label{3:weighted Estimsolutions3}
\int_{\C}|g_h|^2\frac{e^{-{U}^h}}{(1+|z|^2)^2}dx\le
C\int_{\C\setminus\cup\Db(z_k,\d)}|\bar{\partial}\ff|^2{e^{-\P_1}}dx=K
\end{equation}
with a constant $C$ depending on $n,\d,R_n$ but not
depending on $h$. The left-hand side in
\eqref{3:weighted Estimsolutions3} can be  estimated
from below  for large $R$, using \eqref{3:Weighted
estimate1}, which gives
\begin{equation}\label{3:weighted Estimsolutions4}
\int\limits_{R\le |z|\le 2R} |g_h|^2e^{-\P_1}dx\le
(1+R^2)^2K,
\end{equation}
as well as
\begin{equation}\label{3:Weighted Estimate5}
\int\limits_{|z|\ge
2(r_n+\d)}|g_h|^2\frac{e^{-\P_1}}{(1+|z|^2)^2}dx\le
2K,
\end{equation}
so that the  weighted norms of $g_h$ over any annulus
$R<|z|<2R$ are bounded uniformly in $h$ (of course,
the bound may depend on $R$). Recalling now that $\ff$
is a smooth function with compact support, we deduce
from  \eqref{3:weighted Estimsolutions4} that  the
weighted $L_2$ norms of $g_h-\ff$ over the annuli are
bounded uniformly in $h$ as well.

We set $f_h=g_h-\ff$,
$\bar{\partial}f_h=\bar{\partial}g_h-\bar{\partial}\ff=0$,
thus $f_h$ is an entire function, moreover,
\begin{equation}\label{3:normality}
    \int\limits_{R\le|z|\le2R}|f_h|dx\le \left(\int\limits_{R\le|z|\le2R}|f_h|^2e^{-\P_1}dx\right)^{1/2}
    \left(\int\limits_{R\le|z|\le2R}e^{\P_1}dx\right)^{1/2}\le
    C(R)K
\end{equation}
It follows from \eqref{3:normality} that in any
annulus $R<|z|<2R$ the family of entire functions
$\{f_h\}, 0<h<\d,$   has bounded $L_1$-norms,
therefore it has bounded $C^N$-norms of any order $N$
in a smaller annulus (see Theorem 1.2.4 in
\cite{Hormander}) and thus, by maximum principle,
bounded $C^N$-norms in any disk $|z|<R$. Therefore, by
Montel's theorem (see, e.g., \cite{Hille}, Theorem
15.2.5,) this family is compact with respect to
uniform convergence on compacts: there exists a
sequence $h_l\to0$ and an entire function $f$ such
that $f_{h_l}$ converges to $f$ uniformly on any
compact. This implies that the sequence of functions
$g_{h_l}$ converges to $g=f-\ff$ uniformly on any
compact. But now note that for a fixed $k$, the
potential $U_k'(z)$
 equals
  $$U_k'(z)= \frac{10}{\pi}\ln h +5(|z-z_k|^2-h^2)\h^2$$ inside the disk
 $\Db(z_k,h)$,
 while all the other terms in ${U^h}$, see
 \eqref{3:tildeU},
 do not depend on $h$ or are uniformly bounded in $h$.
 Therefore $\exp(-U^h(z))$ has the order
 $h^{-{10}/{\pi}}$ in  $\Db(z_k,h)$,
 and since the
 constant in \eqref{3:weighted Estimsolutions3} is  independent of $h$, the
 sequence $g_{h_l}(z_k)$ may only have $0$ as its limit value. So,
 $g(z_k)=0$ and $f(z_k)=g(z_k)+\ff(z_k)$
 equals $0$ for $k<n$ and $1$ for $k=n$.
 This function $f$ is the one we are looking for,
 because
   $
   \int_{\C}|f|^2e^{-\P}dx=\int_{\C}|f|^2e^{-\P_1-\P_0}dx,$
and the finiteness of the latter integral follows from
the estimate
$$e^{-\P_0(z)}\le C\exp\left(-\frac1{2\pi}\int\D\P_0dx\ln|z|\right)\le
 C\exp(-13/\pi\ln|z|)<C|z|^{-4}$$
for large $|z|$, so that $\int_{\C}|f|^2e^{-\P}dx\le
C\int|f|^2e^{-\P_1}(1+|z|^2)^{-2}dx,$ which is finite
due to \eqref{3:Weighted Estimate5}. So we have found
the function $f_n$. The functions $f_k$, $k<n$ are
constructed in the same way, just the function $\ff$
has to be chosen to be equal $1$ in the disk
$\Dc(z_k,\d)$ and vanishing in $\Dc(z_{k'},\d), k'\ne
k$.
\end{proof}

Having established Theorems~\ref{3:TheoSubhar} and
\ref{3:TheoPos} for a non-negative measure, we have as
our next goal extending the results to measures having
a negative part. The general requirement here is that
the negative part $\mu_-$ is in a certain, each time
concretely defined, sense  weaker than the positive
part $\mu_+$. For the rest of the section we suppose
that the measure $\mu_+$ has infinite flux,
$\mu_+(\R^2)=\mu_+(\C)=\infty$.

Let us first discuss what may be the obstacle for an
entire  function $f$ with finite
$\int|f|^2e^{-2\P_+}dx$ to be quadratic summable with
the weight $e^{-2\P_++2\P_-}$, where $\P_-$ is a
potential for $\mu_-$. It may turn out that $\Psi_-$
grows at infinity, so a certain extra decay of $f$ is
required. On the other hand, the  local singularities
of the potential $\P_-$ can only be  negative, and
they would not cause any trouble since the
introduction of the weight $e^{2\P_-}$ can only
improve the convergence of the integral of $|f|^2$.

So, the easiest result in this direction concerns the
case when we can explicitly estimate the growth of
$\Psi_+$ and then take care of the corresponding term
in the weight.
\begin{corollary}\label{3:CorS1} Suppose that $\mu_-$
has compact support. Then the statements of
Theorems~\ref{3:TheoSubhar} and \ref{3:TheoPos} hold
for $\mu=\mu_+-\mu_-$. \end{corollary}

\begin{proof} Let $\mu_-(\C)=2\pi \Phi, \Phi>0.$ Then
the logarithmic potential $\P_-(z)$ of $\mu_-$ grows
at infinity as $\Phi\ln|z|$. Let $N$ be some integer
larger than $\Phi$. Take $N$ points
$z_1,z_2,\ldots,z_N$ such that each $z_k$  is {\em
not} a common zero for  the space $\Lcc$ of entire
functions $f$ with finite $\int|f|^2e^{-2\P_+}dx$. The
latter can, surely, be achieved, and this, in
particular, means that $e^{-2\P_+}$ belongs to $L_1$
near $z_k$. The conditions $f(z_k)=0,\; k=1,\ldots,N$
define a  subspace $\Lcc_N$ of co-dimension $N$ in
$\Lcc$, so $\Lcc_N$ is infinite-dimensional. Now fix a
polynomial $p(z)$ having simple zeros at the points
$z_k$. The polynomial grows as $|z|^N$ at infinity,
therefore all functions of the form $g =p(z)^{-1}f,\;
f\in \Lcc_N$ are entire and have finite integral
$\int|g|^2e^{-2\P_++2\P_-}dx$.\end{proof}

Relaxing the condition of the compactness of
$\supp\mu_-$, we suppose only that
$\mu_-(\C)=2\pi\Phi$ is finite.

\begin{corollary}\label{3:CorrS2} Suppose that
$\mu_-(\C)<\infty$. Then the statements of of
Theorems~\ref{3:TheoSubhar} and \ref{3:TheoPos} hold
for $\mu=\mu_+-\mu_-$. \end{corollary}
\begin{proof} We choose the potential $\P_-$ for the
measure $\mu_-$ in the form:
\begin{equation}\label{3:PotP-}
\Psi_-(z)=\frac{1}{2\pi}\int_{\Db(0,5)}\ln{|z-w|}d\mu_-(w)+
\frac{1}{2\pi}\int_{\C\setminus\Db(0,5)}\ln\frac{2|z-w|}{
|w|}d\mu_-(w)=U_0(z)+U_1(z),
\end{equation}
   For $|z|>5$ we split $U_1$
into two terms, $U_1=U'+U{''}$, where the first term
corresponds to integration over the disk $|w-z|\le\
|z|/4$ and the second one to the integration over the
rest of the plane. In the disk the expression
$\frac{2|z-w|}{|w|}$  is smaller than 1, the integrand
in \eqref{3:PotP-} is negative, and $U'<0.$ To
estimate  the second term we note that $\frac{2|z-w|}{
|w|}\in [2/3,6]$ for $|w|\ge2|z|$, and $\frac{2|z-w|}{
|w|} \in[\frac{1}{2|z|},\frac35]$ for
$5\le|w|\le2|z|$. Thus the integrand in $U''$ is
majorated by $C_1+C_2\ln|z|$ and therefore
$|U''(z)|\le (C_1+C_2\ln|z|)\mu(\{|z|\ge5\})$. A
similar logarithmic estimate, with coefficient
$\mu(\Db(0,5))$, holds for $U_0$ for large $|z|$.
Thus, as a whole, we have $\Psi_-(z)\le C\log|z|$, and
the proof is concluded exactly as the one for
Corollary~\ref{3:CorS1}.\end{proof}

If the negative part of the measure $\mu $ is
infinite, infiniteness of zero modes can still be
established supposing that $\mu$ becomes non-negative
after an averaging, however we need some additional
local regularity conditions.

Further on consider the following conditions for the
signed measure $\mu=\mu_+-\mu_-$, $\mu_\pm\ge0.$

\begin{condition}\label{7:ConditionA}
There exist constants $r_0>0$ and $\theta_0\in(0,1)$
such that $\mu_+(\Db(z,r_0))\le 2\pi\theta_0$ for any
disk in $\R^2=\C$ with radius $r_0$.\end{condition}

Note that Condition~\ref{7:ConditionA} implies that if
AB solenoids are present, their intensities lie in the
interval $(0,\theta_0)$ for $\Pc_{\max}$.

In particular cases we also  suppose that $\mu_+$
and/or $\mu_-$ satisfy

\begin{condition}\label{7:ConditionC}  There is a constant $A_1$ and a radius $R_1$
 such that
for any disk $\Db(z,R_1)$
 \begin{equation}\label{7:ConditionC:inequality}
    \int\limits_{\Db(z,R_1)}|\ln|z-w||d\mu_\pm(w)\le A_1.
\end{equation}
\end{condition}

In particular, Condition~\ref{7:ConditionC} is
satisfied if the measure $\mu_\pm$ is absolutely
continuous with respect to the Lebesgue measure $dx$
and the corresponding densities  belongs uniformly to
$L_{p,\loc}$ for some $p>1$. Note also that if this
condition is fulfilled for some $R_1$ it holds for any
other $R_1$, with a different constant $A_1$.

For a measure $\mu$ satisfying
Condition~\ref{7:ConditionA}, we consider a potential
$\Psi(z)$, a solution of the equation
   $ \Delta \Psi=\mu,$
as well as the potentials of the measures $\mu_\pm$,
  $  \Delta \Psi_\pm=\mu_\pm, \;\Psi=\Psi_+-\Psi_-.$
The potential $\Psi$ (as well as
$\Psi_\pm$) is determined not uniquely but up to an
arbitrary harmonic function.

The first elementary fact we establish concerns
measures satisfying Condition~\ref{7:ConditionC}. Let
$\chi$ be a smooth non-negative function
 $\chi\in C_0^\infty(\Db(0,R)), \chi=\chi(|z|)$
 for some $R$, such that $\int_{\Db(0,R)} \chi=1$, and we set
$\Psi_R=\Psi*\chi$, so that
$\Delta\Psi_R=\mu_R=\mu*\chi$. We set also
$\Psi_{\pm,R}=\Psi_\pm*\chi$,
$\Delta\Psi_{\pm,R}=\mu_{\pm,R}\equiv\mu_\pm*\chi$.
\begin{lemma}\label{7:SimpleLemma} Suppose that the measure
$\mu_-$ or $\mu_+$ satisfies
Condition~\ref{7:ConditionC}. Then there  is a
constant $C=C(R,R_1,A_1)$ such that, with the
corresponding sign $\pm$,
\begin{equation}\label{7:SimpleLemma:Formula}
    |\Psi_\pm(z)-\Psi_\pm^*(z)|\le C.
\end{equation}
\end{lemma}
\begin{proof} Denote by $\Db$ the disk $\Db(z,R)$ and
by $\Db'$ the concentric disk with twice as large
radius. Split the measure $\mu_+$ into the sum of the
measure $\mu_{+}'$ supported in  $\Db'$ and
$\tilde{\mu}_{+}$ supported outside this disk.
Correspondingly, the potential $\Psi_+$ splits into
the sum of $\Psi_+'=\mu_+'*G_0$ and
$\tilde\Psi_+=\Psi_+-\Psi_+'$,
$G_0(z)=(2\pi)^{-1}\ln|z|$. The function
$\tilde\Psi_+$ is harmonic in $\Db'$, therefore
$\tilde\Psi_+*\chi=\tilde\Psi_+$ in the disk $\Db$, in
particular, at the point $z$. The potentials $\Psi_+'$
and $\Psi_+'*\chi$ are bounded, by
\eqref{7:ConditionC:inequality}. This proves the
Lemma.\end{proof}
 Next we will study the potentials of
measures for which  $\mu_+$ satisfies
Condition~\ref{7:ConditionA}, and  another component,
$\mu_-$ satisfies Condition~\ref{7:ConditionC}. Fix
some  $R$. For a fixed point $z_0\in\C$ denote by
$\Db_0, \Db, \Db_1, \Db_2$ the disks with center at
$z_0$ and radii, respectively, $r_0/2,r_0, r_0+R,
r_0+2R $. We fix a non-negative mollifier $\chi\in
C_0^\infty(\Db(0,R))$ as above and set
   $ \Psi_R(z) = \chi * \Psi,\;
   \Psi_{\pm,R}=\chi*\Psi_\pm.$

We prove now our main local estimate.
\begin{proposition}\label{7:Prop:ineq}
 For a fixed $z$, suppose that, Condition~\ref{7:ConditionA}
  is satisfied for $\mu_+$, and   Condition~\ref{7:ConditionC} is satisfied for $\mu_-$
  in $4R$-neighborhood of $z$.
  Then there exist  constants $C_0=C_0(r_0,R, \theta_0, A_1)$,
    $C_1=C_1(r_0,R, \theta_0, A_1)$, and  $C_2=C_2(r_0,R, \theta_0, A_1)$ such
that
\begin{equation}\label{7:IneqDisk}
    \int\limits_{\Db_0}e^{-2\Psi(z)}|f(z)|^2dx\le
    C_0\int\limits_{\Db_1}e^{-2\Psi_R(z)}|f(z)|^2dx+
    C_1\int\limits_{\Db_1}e^{-2\Psi_R(z)}|\partial f(z)|^2dx
\end{equation}
and
\begin{equation}\label{7:IneqDiskPM}
   \int\limits_{\Db_1}
  e^{2\Psi_{-,R}(z)}|f(z)|^2dx\le
C_2 \int\limits_{\Db_1}e^{2\Psi_{-}(z)}|f(z)|^2dx
\end{equation}
for any function $f\in L_{2}(\Db_1)$, as soon as the
inequalities make sense. The derivative $\partial$ in
\eqref{7:IneqDisk} can be replaced by
$\tilde{\partial}$.
\end{proposition}
\begin{proof} For brevity, we prove the inequalities
 for the disks centered at
the origin,
 noticing that the constants in all estimates below depend
 only on $r_0,R, \theta_0,A_1$. First, due to
 Lemma~\ref{7:SimpleLemma}, we can restrict ourselves
 to a non-negative measure $\mu$ since the negative
 part of $\mu$ contributes to the estimates only with a constant factor
 when passing from $e^{-\Psi}$ to $e^{-\Psi_R}$.

We split the measure $\mu$ into the sum
 $\mu=\mu'+{\nu}$, so that $\mu'$ is supported in
 the disk
  $\Db_2$   and ${\nu}$ is supported
 outside this disk. The potential
 $\Psi$ splits into two parts,
 $\Psi=\Psi'+H$, where
   $ \Psi'=G*\mu'$
is the Newton potential of the measure $\mu'$,
$G(z)=1/(2\pi)\ln\left|{z}/{R}\right|,$ and
$H(z)=\Psi(z)-\Psi'(z)$ is a harmonic function inside
the disk $\Db_2$.

Correspondingly, the smoothened potential $\Psi_R$
splits into two terms,
  $  \Psi_R=\Psi_R'+H_R,$
where $\Psi_R'=\Psi'*\chi$, $H_R=H*\chi$.  Note that
since  $H$ is harmonic inside $\Db_2$, the functions
$H$ and $H_R$ coincide inside $\Db_1$. The function
$\Psi_R'=\mu'*G*\chi$ is bounded in $\Db_1$,
$|\Psi_R'|\le c_1=c_1(r_0,R,A_1,\theta_0)$.

Let $g(z)$ be a function, anti-analytical in $\Db_1$,
such that $H(z)=-\ln(|g(z)|)$. Then we have for any
$f$
\begin{eqnarray*}\label{7:TransformedTerms}
\int_{\Db_0}e^{-2\Psi(z)}|f(z)|^2dx=
\int_{\Db_0}e^{-2\Psi'(z)}|f(z)g(z)|^2dx,\\
\int_{\Db_1}e^{-2\Psi_R(z)}|f(z)|^2dx=
\int_{\Db_1}e^{-2\Psi_R'(z)}|f(z)g(z)|^2dx,\\
\int_{\Db_1}e^{-2\Psi_R(z)}|\partial
f(z)|^2dx=\int_{\Db_1}e^{-2\Psi_R'(z)}|{\partial}(f(z)g(z))|^2dx.
\end{eqnarray*}
So,  denoting $u=fg$ and taking into account the
boundedness of  $\Psi_R'(z)$, we see that it is
sufficient to establish the estimate
\begin{equation}\label{7:FinalTransfInDisk}
    \int\limits_{\Db_0}e^{-2\Psi'(z)}|u|^2dx\le
    C\int\limits_{\Db_1}(|u|^2+|{\partial}u|^2)dx
\end{equation}
To prove \eqref{7:FinalTransfInDisk}, we split $\mu'$
into further  two parts,
  $  \mu'=\mu_0+\mu_1$
where $\mu_0$ is supported in $\Db$ and $\mu_1$ in
$\Db_1\setminus\Db$. The function $\Psi_1=\mu_1*G$ is
bounded in $\Db_0$. In fact, the distance between
points in $\Db_0$ and in the support of $\mu_2$ lies
between $r_0/2$ and $R+2r_0$, therefore
$$|\Psi_1(z)|\le \max(|\ln (r_0/(2R)|,|\ln(R+2r_0)/R|)|\mu|(\Db_1)\le C.$$

If $\mu_0=0$, the required inequality is now obvious.
Otherwise, in order  to estimate the contribution of
$\Psi_{0}=\mu_0*G$, we apply Jensen's inequality:
  $$ e^{-2\mu({\Db})\int_\Db
    G(z-w)\frac{d\mu(w)}{\mu({\Db})}}\le\nonumber\\
    \int_\Db e^{-2\mu({\Db})
    G(z-w)}\frac{d\mu(w)}{\mu({\Db})}=R^{2\mu({\Db})}\int_\Db
    |z-w|^{-\mu({\Db})/\pi}\frac{d\mu(w)}{\mu({\Db})}.$$
So, for the left-hand side in
\eqref{7:FinalTransfInDisk} we have
\begin{equation}
\label{7:StartEstimatingInDisk}
   \int_{\Db_0}e^{-2\Psi_{0}(z)}|u(z)|^2dx\le
   C\int_{\Db}\left[ \int_{\Db_0}|u(z)|^2
   |z-w|^{-\mu({\Db})/\pi}dx\right]\frac{d\mu(w)}{\mu({\Db})}.
   \end{equation}
In the inner integral we apply the H\"older
inequality, taking into account that
$\mu({\Db})/\pi<2\theta_0<2$:
\begin{equation}\label{7:Holder}
 \int_{\Db_0}|u(z)|^2
   |z-w|^{-\mu({\Db})/\pi}dx\le
   ||u||_{L_{2q}(\Db_0)}^2||
   |z-w|^{-\mu({\Db})/\pi}||_{L_{q'}(\Db_0)}\le
   C||u||_{L_{2q}(\Db_0)}^2,
\end{equation}
provided $q\in(1,\infty))$ is chosen so that
$\theta_0q'<1$, therefore  the norm of
$|z-w|^{-\mu({\Db})/\pi}$ in \eqref{7:Holder} is
finite. The second integration in
\eqref{7:StartEstimatingInDisk} gives then

\begin{equation}\label{7:Holder+1}\int\limits_{\Db_0}e^{-2\Psi_{0}(z)}|u(z)|^2dx\le
C ||u||_{L_{2q}(\Db_0)}^2.\end{equation} Finally we
apply the Sobolev type embedding theorem in the disk
$\Db_0$:
$$||u||_{L_{2q}(\Db_0)}^2\le
C(||u||_{L_2(\Db_1)}^2+||\partial
u||_{L_2(\Db_1)}^2)$$ and recall \eqref{7:Holder+1}
and Lemma~\ref{7:SimpleLemma}. This proves
\eqref{7:FinalTransfInDisk}, and therefore
\eqref{7:IneqDisk}. The inequality
\eqref{7:IneqDiskPM} follows immediately from
Lemma~\ref{7:SimpleLemma}. Obvious changes establish
Lemma for $\bar{\partial}$.
\end{proof}
De-localizing  \eqref{7:IneqDisk},
\eqref{7:IneqDiskPM} leads to the following
fundamental fact.
\begin{proposition}\label{7:PropFundamentalInequality}
Suppose that for all $z$ the measure $\mu_+$ satisfies
Condition~\ref{7:ConditionA}, and $\mu_-$ satisfies
Condition~\ref{7:ConditionC},  $\Psi(z)$ is a
potential for the measure $\mu$ and $\Psi_R$ is the
smoothened potential $\Psi_R=\Psi*\chi$. Then, with
some constants $C_0, C_1, C_2$ depending only on $r_0,
R_1, \theta_0, A_1$
\begin{equation}\label{7:FundamentalInequality}
    \int\limits_{\C}e^{-2\Psi(z)}|f(z)|^2dx\le
    C_0\int\limits_{\C}e^{-2\Psi_R(z)}|f(z)|^2dx+
    C_1\int\limits_{\C}e^{-2\Psi_R(z)}|{\partial}f(z)|^2dx,
\end{equation}
and
\begin{equation}\label{7:FundamentalInequalityPM1}
\int\limits_{\C}e^{-2\Psi_R(z)}|\partial f(z)|^2dx\le
C_2\int\limits_{\C}e^{-2\Psi(z)}|\partial f(z)|^2dx,
\end{equation}
 as soon as the inequalities \eqref{7:FundamentalInequality}, resp.,
  \eqref{7:FundamentalInequalityPM1}, make sense.
  Again, $\partial$ can be replaced by
  $\bar{\partial}$.
\end{proposition}
\begin{proof}

We take a covering of $\C$ by the disks $\Db$ with
radius $r_0/2$ such that the concentric disks with
radius $R+r_0/2$ form a covering with finite
multiplicity $\varkappa$. Then we write the estimate
\eqref{7:IneqDisk} for each disk $\Db$, and sum these
inequalities. This leads us to
\eqref{7:FundamentalInequality}. To prove
\eqref{7:FundamentalInequalityPM1}, consider the usual
splitting $\Psi=\Psi_+-\Psi_-$. The function $\Psi_+$
is subharmonic, therefore $\exp(-2\Psi_{+,R})\le
\exp(-2\Psi_+)$, so it is sufficient   to establish
$$\int\limits_{\C}e^{2\Psi_{-,R}(z)}|h(z)|^2dx\le
C_2\int\limits_{\C}e^{2\Psi_-(z)}|h(z)|^2dx,$$ where
$h(z)=\exp(-\Psi_+)|\bar{\partial}f(z)|$. The latter
inequality follows immediately from its localized
version \eqref{7:IneqDiskPM}.
\end{proof}

\begin{proposition}\label{3:opposite ineq} Suppose that the measure $\mu_-$
satisfies Condition~\ref{7:ConditionC} Then
\begin{equation}\label{7:FundamentalIneqPMdeloc}
 \int\limits_{\C}e^{-2\Psi_R(z)}|
h(z)|^2dx\le C_2\int\limits_{\C}e^{-2\Psi(z)}|
h(z)|^2dx,
\end{equation}
 as soon as
  \eqref{7:FundamentalIneqPMdeloc}, makes sense.
\end{proposition}
\begin{proof}
To prove \eqref{7:FundamentalIneqPMdeloc}, note that
the function $\Psi_+$ is subharmonic, therefore\\
$\exp(-2\Psi_{+,R}) \le \exp(-2\Psi_+)$, so it is
sufficient   to establish
$$\int\limits_{\C}e^{2\Psi_{-,R}(z)}|f(z)|^2dx\le
C_2\int\limits_{\C}e^{2\Psi_-(z)}|f(z)|^2dx,$$ where
$f(z)=\exp(-\Psi_+)|h(z)|$. The latter inequality
follows immediately from Lemma~\ref{7:SimpleLemma}.
\end{proof}

Now we can establish our next  theorem on zero modes.
We recall that for the operator $\Pc_{max}$ we accept
normalization of the discrete part of the measure
$\mu$ so that all intensities of AB-solenoids lie in
$(0,1)$, while for the operator $\Pc_{EV}$ these
intensities lie in $[-\frac{1}{2},\frac{1}{2})$, with
$0$ excluded.
\begin{theorem}\label{7:MainTheorem}
Suppose that the measure $\mu_-$ satisfies
Condition~\ref{7:ConditionC},  $\mu$ satisfies
Condition~\ref{7:ConditionA}, moreover, for a certain
$R_0>0\;\;,$ $A(z)=\mu(\Db(z,R_0))\ge0$ for $|z|$
large enough and $\int_{\C}A(z)dx=\infty.$
 Then
the spin-down components of the Pauli operators
$\Pc_{-\max}$, $\Pc_{-EV}$ have an
infinite-dimensional null subspace. If, additionally
 a stronger
condition
\begin{equation}\label{3:ShigekCondStrong}
A(z)\ge A_0>0
\end{equation}
is satisfied for all $|z| $ large enough then the
point zero is an isolated point in the spectrum of
$\Pc_{-EV}$, $\Pc_{-\max}$ and the spectral gap above
zero is estimated from below by some constant
depending on $r_0, R_0, R_1 , \theta_0, A_0,
A_1$.\end{theorem}

\begin{proof}  Note first that due to our normalization agreement,
see Sect.2, the null subspace in both $EV$ and $\max$
cases is generated by entire functions. Therefore we
suppress the corresponding subscript. Let $\Psi$ be a
potential for $\mu$. The null subspace of the operator
$\Pc_-$ consists of the functions of the form $u\in
L_2(\C)$ such that $u=\exp(-\Psi)f$, with an entire
(anti-analytical) function $f(z)$. Take some $R>R_0$
and a mollifier $\chi_0$ supported in the disk
$\Db(0,R-R_0)$. Set $\chi=\chi_0*\chi_{R_0}$, where
$\chi_{R_0}$ is the characteristic function of
$\Db(0,R_0)$. Under the conditions of the first part
of the theorem, the potential $\Psi_R(z)=\Psi*\chi$ is
subharmonic outside a compact set.
 If $f$ is an entire function, the
second term on the right-hand side in
\eqref{7:FundamentalInequality} vanishes, so
 we obtain the estimate
\begin{equation}\label{6:ImportantEstimate}
    \int\limits_\C e^{-2\Psi(z)}|f(z)|^2dx\le
    C_0\int\limits_\C e^{-2\Psi_R(z)}|f(z)|^2dx.
\end{equation}
This inequality implies that if for some entire
function $f$ the function $\exp(-\Psi_R)f$ belongs to
$L_2$ then $\exp(-\Psi)f$ also belongs to $L_2$.
However the space of the functions satisfying the
former condition is infinite-dimensional by
Corollary~\ref{3:CorS1}. Therefore the space of entire
functions $f$ with $\exp(-\Psi)f\in L_2$ is also
infinite-dimensional.

To prove the existence of the spectral gap,  we note
first of all that under the conditions of the second
part of the theorem we can, by increasing $R_0$, have
$A(z)>A_0/2$ for all $z$, so, that $\Psi_R$ is
strictly subharmonic.  From
  H\"ormander's theorem (Lemma~4.4.1 in
  \cite{Hormander}) on solutions of the
$\bar{\partial}$-equation in weighted spaces  (we
apply this Lemma with $\partial$ replacing
$\bar{\partial}$), it follows that for any function
$g\in L_2$ such that $\exp(-\Psi_R)g\in L_2$, there
exists a solution $f$ of the equation ${\partial}f=g$
such that
\begin{equation}\label{7:HormanderEstimate}
\int\limits_\C e^{-2\Psi_R(z)}|f(z)|^2dx\le
c\int\limits_\C e^{-2\Psi_R(z)}|g(z)|^2dx.
\end{equation}
Substituting this inequality into
\eqref{7:FundamentalInequality}, and using
\eqref{7:FundamentalIneqPMdeloc}, we obtain
\begin{equation}\label{7:GapIneq1}
    \int_\C
    e^{-2\Psi(z)}|f(z)|^2dx\le c' \int_\C
    e^{-2\Psi(z)}|g(z)|^2dx.
\end{equation}
We set $e^{-\Psi(z)}f(z)=u(z), e^{-\Psi(z)}g(z)=v(z),
$ and recalling that $g={\partial}f$ and
$Q_-=e^{-\Psi}{\partial}e^{\Psi}$ we get the
inequality
\begin{equation}\label{7:GapIneq2}
\|u\|_{L_2}^2\le C\|v\|_{L_2}^2, v=Q_-u,
\end{equation}
where $v$ is an arbitrary function in $L_2$ and $u$ is
a certain function in $L_2$ satisfying $Q_- u=v$. Let
$h(z)$ be the projection of $u$ onto the subspace
$\Nc$ in $L_2$ orthogonal to all solutions $w$ of the
equation $Q_-w=0$; we still have $v=Q_-h$. The
left-hand side of \eqref{7:GapIneq2} can only decrease
if we replace there $u$ by $h$. So, on the subspace
$\Nc$ which is the spectral subspace of the Pauli
operator $\Pc_{-} =Q_{-}^*Q_{-}$, corresponding to the
nonzero spectrum, the norm of the function $h$ is
majorated by $\pF_-[h]$, the value of the quadratic
form of the operator $\Pc_-$ on the function $h$. This
exactly means that the nonzero spectrum of $\Pc_{-}$
is separated from zero, and the width of the spectral
gap is controlled by the constant $C$ in
\eqref{7:GapIneq2}.
\end{proof}

The restrictions imposed in
Theorem~\ref{7:MainTheorem} on the negative part of
$\mu$ can be relaxed.
\begin{corollary}\label{3:Corollary} Suppose that the
measure $\mu$ satisfies conditions of the first part
of Theorem~\ref{7:MainTheorem} and $\nu$ is a finite
non-negative measure. Then for the measure $\mu-\nu$
the zero energy subspace is
infinite-dimensional.\end{corollary}

The Corollary follows immediately from the main
theorem and Corollary~\ref{3:CorrS2}.

 If there are arbitrarily large
regions   where the field vanishes, we cannot use the
second part of Theorem~\ref{7:MainTheorem} to
establish the presence of the spectral gap, even for a
non-negative measure since such regions violate the
condition \eqref{3:ShigekCondStrong}.   The following
statement shows that if there are such regions then,
actually, zero cannot be an isolated point in the
spectrum. We suspect that the result is known to
specialists, but we could not find a reference, so we
present a proof, for the sake of completeness.
\begin{proposition}\label{remark on gaps} Suppose that
for any $R>0$ there exists a disk $\Db$ with radius
$R$ where the measure $\mu$ is zero. Then zero is not
an isolated point in the spectrum of the Pauli
operators $\Pc_-$ and $\Pc_+$.\end{proposition}
Note
that we do not suppose anything about the nature of
zero as a point in the spectrum.

\begin{proof}
To justify the  statement, we find, for any $\e>0$, a
function $\phi$ orthogonal to the null subspace of the
operator $\Pc_-$ (or $\Pc_+$) such that $\pF_-[\phi]$,
resp., $\pF_+[\phi]$ is smaller than $\e||\phi||^2$.
To do this (for $\Pc_-$, for example), we fix a
non-trivial smooth function $u_0\ge 0$, with compact
support in the unit disk $\Db_1$. For some constant
$C$, the estimate
    $||u_0||_{L_2}\le C||\nabla u_0||_{L_2}$
holds.
 Now, let $\Db$ be a disk with
radius $R>\e^{-1}$ and center at $z_0$, such that the
restriction of $\mu$ to $\Db$ is zero, and $u$ be the
function in $C^\infty_0$ obtained from $u_0$ by the
dilation and shift, $u(z)=u_0(R^{-1}(z-z_0))$. The
function $u$ satisfies
\begin{equation}\label{4:estimateForProjection2}
||u||_{L_2}\le CR^{-2}||\nabla u||_{L_2}
\end{equation}

Since the magnetic field is regular (in fact, it zero)
on the support of  $u$, the function $u$ belongs to
the domain of the operator $\Pc_-$

 Denote by $v$ the projection
of $u$ to the null space of the operator $\Pc_-$. The
function $v$ also belongs to the domain of the
operator $\Dc_-$ (the function $v$ is zero if, in
particular, the null space is trivial.) So, $\phi=u-v$
also belongs to the domain of $\Pc_-$ and is
orthogonal to $\Ker(\Pc_-)$; $\phi$ is nontrivial
since $v$, being a zero mode, cannot have compact
support (unless $v=0$). The norm of the function
$\phi$ in $L_2(\R^2)$ is not greater than the norm of
$u$ in $L_2(\R^2)$, or, what is the same, than
$||u||_{L_2(\Db)}$. At the same time, for the
quadratic form of the operator $\Pc_-$, we have
   $$ \pF_-[\phi]=\pF_-[u]=
    \int_{\Db}\exp(-2\Psi)|\partial_z(\exp(\Psi)u)|^2dx.$$
However the magnetic field $\mu=\Delta \Psi$ vanishes
in $\Db$, so $\Psi$ is a harmonic function in $\Db$.
Taking into account that $u$ has its support in $\Db$,
by means of the usual partial integration we obtain
that $\pF[u]=\int_\Db|\nabla u|^2 dx.$ Now, the
inequality \eqref{4:estimateForProjection2} gives us
\begin{equation}\label{4:FinalOnProjection}
    ||\phi||^2_{L_2(\R^2)}\le cR^{-2}\pF_-[\phi],
\end{equation}
for the function $\phi$ orthogonal to the null
subspace of $\Pc$. Supposing that there are disks of
arbitrary size $R$, not intersecting the support of
$\mu$ we obtain that in the neighborhood of the point
zero there are infinitely many points of the spectrum
of $\Pc_-$.\end{proof}
 The reasoning in the proof of
Theorem~\ref{7:MainTheorem} does not apply directly to
the operator $\Pc_{+\max}$, if, for example, the
measure $\mu$ is continuous. In fact, when we pass to
$\Pc_{+}$, we have to replace $\mu$ by $-\mu$ and
${\partial}$ by $\bar{\partial}$. The latter change is
not that essential. However the measure $-\mu$ does
not satisfy  the condition $-\mu(\Db(z,R_0))\ge0$,
this quantity is negative and the line of reasoning
breaks down in several places.

The case when the game can be saved for the operator
$\Pc_{\max}$ is the one with a measure $\mu$
consisting of the system of Aharonov-Bohm solenoids
placed at the points of a discrete set $\L$, with
intensities $\a_\l,\;\a_\l\in(0,1)$,  on the
background of a continuous field $\nu$ such that both
$\nu_+$ and $\nu_-$ satisfy
Condition~\ref{7:ConditionC}:
\begin{equation}\label{8:AB+nu}
    \mu=\mu_{AB}+\nu,
    \mu_{AB}=2\pi\sum_{\l\in\L}\a_\l\d(z-\l).
\end{equation}
In this case, as  explained in Sect. 2, the measure
$-\mu$ can be reduced by a gauge transform to the one
containing  AB-fluxes  with intensities $(1-\a_\l)$,
i.e., to the measure
\begin{equation}\label{8:AB-nu}
    \tilde{\mu}= \tilde{\mu}_{AB}-\nu,\;\;
    \tilde{\mu}_{AB}=2\pi\sum_{\l\in\L}(1-\a_\l)\d(z-\l).
\end{equation}
If the set $\L$ is infinite, does not have large gaps,
and the numbers $\a_l$ are separated both from $0$ and
$1$, it is possible that  both measures $\mu,
\tilde{\mu}$ satisfy conditions of
Theorem~\ref{7:MainTheorem}. This leads to
infiniteness of zero modes and the spectral gap for
both $\Pc_{+\max}$ and $\Pc_{-\max}$

We formulate a special case, where Conditions on the
measures are expressed in more geometrical terms:

\begin{condition}\label{3:Cond4}
   There exist positive numbers $ r^0,R_0$  such
    that any disc $\Db(z,R_0)$  contains at least one point in
    $\L$,  any disk $\Db(\l,r^0),\;\l\in\L,$ contains no points in $\L,$ other than
    $\l$,
    and all intensities $\a_\l, \;\l\in\L$,
     satisfy $\a_\l\in(\theta_0,1-\theta_0)$ for a
    certain $ \theta_0>0.$
\end{condition}

\begin{corollary}\label{8:Corr:P+AB} Suppose that the
 measure $\m$ has the form \eqref{8:AB+nu}
its  discrete part satisfies Condition~\ref{3:Cond4}.
Suppose also that both positive and negative parts of
 the continuous part $\n$ in \eqref{8:AB+nu}
satisfy Condition~\ref{7:ConditionC}, moreover, for
$|z|$ large enough,
$\theta_0\ge\nu_\pm(\Db(z,R_0)).$
Then both operators $\Pc_{\pm,\max}$ have an
infinite-dimensional null subspace. If, moreover,
$\theta_0\ge\nu_\pm(\Db(z,R_0))\ge A_0>0,$ both
operators possess a spectral gap. The size of the gap
is determined by the numbers $r^0,R_0,\theta_0, A_1$,
and $a_0$.\end{corollary}

The Corollary above covers, among other cases, a
purely discrete measure, i.e., an infinite
configuration of AB solenoids satisfying
Condition~\ref{3:Cond4}, as well as such a
configuration on the background of a constant or
'almost constant' magnetic field. For a regular
lattice $\L$ and equal intensities and a constant
background field, the infiniteness of zero modes was
established in \cite{GeyGri}, \cite{GeySto}.

On the other hand, if the measure $\mu$ is continuous
(thus $\Pc_{\max}=\Pc_{EV}$), so that there are no AB
solenoids, we can establish the spectral gap also for
$\Pc_{+}$.

\begin{corollary}\label{3:Corollary2} Let the measure
$\mu$ satisfy Condition~\ref{7:ConditionA},
$\mu(\Db(z,R_0))\ge A_0>0$ for $|z|$ large enough, and
both $\mu_\pm$ satisfy Condition~\ref{7:ConditionC}.
Then the operator $\Pc_{+}$ has no zero modes and
possesses a spectral gap.\end{corollary}
\begin{proof} Let $\Psi$ be the potential of the measure $\mu$,
 $\Psi_R$ its averaging, so that $\Delta\Psi_R=\mu*\chi.$
 is a measure absolutely continuous with respect to
 Lebesgue measure with positive density separated from
 zero.
As it follows from Proposition~\ref{3:opposite ineq}
and the conditions imposed on $\mu$, for any function
$f$,
\begin{equation}\label{3:norms}
    \int \exp(2\Psi)|f|^2 dx\ge C\int \exp(2\Psi_R)|f|^2
    dx.
\end{equation}
Now  for the smooth function $\Psi_R$ we use the
commutational relation which gives
\begin{equation}\label{3:comrelations}
\int\limits_{\C}|(e^{\Psi_R(z)}u)\partial
(e^{-\Psi_R(z)}u)|^2dx=\int\limits_{\C}e^{-2\Psi_R(z)}|\partial_{\bar{z}}
(e^{\Psi_R(z)}u)|^2dx+4\int\limits_{\C}|u|^2 d\mu_R,
\end{equation}
And this establishes our statement.
\end{proof}

An important special case of the above considerations
concerns  periodic magnetic fields. Such
configurations attracted interest in early 80-s. In
the papers \cite{DubNov80} and \cite{DubNov80a} for
the case of a rational flux of the field over an
elementary cell of the lattice the infiniteness of
zero modes was proved as well as the existence of the
spectral gap, and nothing has been done since. We show
below that the above restriction is irrelevant.

\begin{corollary}\label{3:Ex1} Let the measure $\mu$ be
periodic with respect to some lattice in the plane.
Suppose that  Condition~\ref{7:ConditionC} is
satisfied for both positive and negative parts of the
measure $\mu$ and, moreover, the measure of one cell
$\digamma$ of the lattice is positive. Then there are
infinitely many zero modes for the Pauli operator
$\Pc$ and a spectral gap.
\end{corollary}

 In fact, for a periodic measure, the Conditions
 \ref{7:ConditionA} and positivity of $\mu(\Db(z,R_0))$ for
   large $|z|$ obviously follow from the positivity of
 the measure of the cell, and this, by
 Theorem~\ref{7:MainTheorem}, establishes both
 infiniteness of zero modes and spectral gap for
 $\Pc_-$. The presence of a spectral gap for   $\Pc_+$ follows from
  Corollary~\ref{3:Corollary2}.

For a periodic magnetic field the infiniteness of zero
modes may be proved also in the following way. Let
$\L$ be the lattice of periods of the measure $\mu $,
with the elementary cell $\digamma$. Let $\sigma(z)$
be the Weierstrass function of the lattice $\L$
defined in \eqref{2:Weierstrass}. As it is explained
in Sect.2, with proper $\n$, the function
$\Psi_0=\ln|\sigma(z)|-\Re(\nu z^2)$ equals
$m|z|^2+\r(z)$, $m=\frac{\pi}{2|\digamma|}$ with a
periodic function $\r(z)$ having singularities of the
form $(2\pi)^{-1}\ln|z-\l|$ near each point $\l$ of
the lattice $\L$, $|\digamma|$ being the area of the
elementary cell. Consider the potential
\begin{equation}\label{7:periodic potential}
\Psi_\mu(z)=\int_\digamma \exp(\Psi_0(z-w))d\mu(w)=
\int_\digamma
|z-w|^2d\mu(w)+\int_\digamma\r(z-w)d\mu(w),
\end{equation}
with  integration over the elementary cell $\digamma$
of $\L$. Since $\Delta\Psi_0=\sum\delta(z-\l)$, the
function $\Psi$ is a potential for the measure $\mu$.
From the Condition~\ref{7:ConditionC} for $\mu_\pm$,
it follows that the second term in \eqref{7:periodic
potential} is a periodic bounded function and
positivity of the flux $\Phi$ over the elementary cell
produces the growth of the first term in
\eqref{7:periodic potential} as $\Phi m|z|^2$.
Therefore any function having the form
$u=\exp(-\Psi_\mu(z))f$ with an entire function $f(z)$
growing not faster that $\exp(c|z|^2)$, $c<\Phi m$, is
a zero mode for the periodic magnetic field $\nu$.

More generally, we can consider a quasi-periodic
magnetic field. Let $\mu=\sum\mu_k$, $k=1,\dots, N$,
so that each measure $\mu_k$ is periodic with respect
to its own lattice $\L_k$. Suppose also that
Condition~\ref{7:ConditionC} is satisfied for each
measure $\mu_k$, and the sum $\Phi=\sum_k
\frac{\Phi_k}{|\digamma_k|},$ $|\digamma_k|$ being the
area of the cell of the lattice of periods for
$\mu_k$, is positive. Then the reasoning used for a
single periodic field goes through. Again, the
smoothened field $\mu*\chi$ is a measure with a
bounded positive density, which guarantees,  as in
Corollary~\ref{3:Ex1}, infiniteness of zero modes and
the presence of a spectral gap. Moreover, the
explicitly constructed potential, being the sum of
potentials $\Psi_{\mu_k}$, grows as $c|z|^2$.

Of course  if the sum is negative, then the potential
for the measure $\mu=\sum\m_k$  is majorated by
$-c|z|^2$, and this leads to infinitely many zero
modes for the Pauli operator $\Pc_+$.

We note here that the  fields above can be,
  by Corollary~\ref{3:Corollary}, perturbed by any finite measure
   thus preserving
  the infiniteness of zero modes.

In the following sections we consider more strong
perturbations of the field configurations described
above. These perturbation preserve the quadratic lower
estimate for the potential, thus guaranteeing
infiniteness of zero modes,  however we will see that
they may destroy the spectral gap.

\section{Perturbations of the field and zero modes}

From now on we consider a more special situation than
in the previous section.  Suppose that for a certain
magnetic field $\m_0$ we know that there exists a
potential $\Psi_0$  satisfying the growth condition
$\Psi_0(z)\ge \g |z|^2$  for $|z|$ large enough, with
some $\g>0$. Then any entire analytical function $f$
growing at infinity slower than $\exp(\g-\e)|z|^2$
satisfies $f\exp(-\Psi)\in L_2$, and thus the null
subspace of the Pauli operator $\Pc_-$ is
infinite-dimensional. Of course, this does not
guarantee the presence of the spectral gap, and,
moreover, the spectral gap may fail to exist, as we
show below. Now, let $\m_1$ be another magnetic field
with a potential $\Psi_1(z)$ satisfying the estimate
$\Psi_0(z)-\Psi_1(z)=o(|z|^2)$ or even
$|\Psi_0(z)-\Psi_1(z)|\le \g^*|z|^2$ with $\g^*<\g$,
for large $|z|$. Then, of course, the potential
$\Psi_1$  grows at infinity sufficiently fast so that
the infiniteness of the number of zero modes for the
Pauli operator with field $\mu_1$ is granted. More
generally, the estimate for $\Psi_0$ and the
inequality for the difference $\Psi_0(z)-\Psi_1(z)$
may contain some singular terms, as it happens in the
presence of A-B solenoids.

There are several types of the magnetic fields that we
can take as the unperturbed field $\mu_0$. One obvious
example is the constant magnetic field described by
the measure $d\mu_0=Bdx$, proportional to the Lebesgue
measure -- here the potential has the form
$\Psi_0=\frac{B}{4}|z|^2$. As  shown in the end of the
previous section, a more general field, a periodic
field with mild local regularity properties (Condition
~\ref{7:ConditionC}) with nonzero flux through the
cell, or the even more general one, a quasi-periodic
field with nonzero $\sum_k
\frac{\Phi_k}{|\digamma_k|}$, also possess  potentials
subject to required growth conditions.

Another kind of the starting point of our study can
the observation made in \cite{GeyGri}. Let $\L=\L_0$
be a regular, periodic lattice with periods
$\omega_1,\omega_2$ and all intensities are equal,
$\a_\l=\a\in(0,1),\, \l\in\L$. The scalar potential
$\Psi_{0+}(z)$  is defined in
\eqref{2:realperiodicity}.  Since $\alpha$ lies
between $0$ and $1$, the function
$u(z)=f(z)\exp(-\Psi_{0+}(z))=f(z)\exp(-\a
m|z|^2)\exp(-\a\r(z))$ belongs to $L_2$ as soon as the
entire function $v(z)$ grows at infinity not faster
than $\exp{\g'|z|^2}$, $\g'<\a m$. Of course, there
are a lot of such entire functions $v(z)$, in
particular, all polynomials fit. This proves the
infiniteness of zero modes for $\Pc_{+\max}$ or, under
the condition $\a\in(0,\frac12)$, for $\Pc_{+EV}$.
Taking into account the possibility of  applying the
gauge transformations discussed in the end of Sect. 2,
the same reasoning applies to the operator
$\Pc_{-\max}$.  Again we can perturb the (discrete
now) measure $\mu_0$ by a measure having a potential
with a slower growth than $\Psi_0$ and with controlled
logarithmic singularities.

So, in order to determine which perturbations of the
initial field $\mu_0$ preserve infiniteness of zero
modes, one has to know which measures possess
potentials with prescribed control over the behavior
at infinity and, if needed, at singular points. A
number of such results exist in the literature, see,
e.g., \cite{HaymanKenn}, however they are not
sufficient in our situation since they require an
extra  decay of the measure at infinity, the condition
we  aim to avoid. Moreover, they do not usually take
into account the possible cancellation of the
contribution of the positive and negative parts of the
perturbing measure. In this section we present some
results on the estimates for potentials for certain
classes of measures.

The situation we consider first is the one when the
whole (or a part of) measure $\mu_0$ is re-arranged,
more exactly, this measure is replaced by its image
under some mapping of the plane. Under such
re-arrangement, large regions  with  field having
'wrong direction' may arise, so that the positivity
conditions of the general theorems are substantially
broken. Nevertheless, the infiniteness of zero modes
is preserved. We consider the case of $\mu$ being a
continuous measure with $\mu_\pm$ satisfying
 Condition~\ref{7:ConditionC}. Let $\F:\R^2\to\R^2$ be a Borel measurable mapping
 and $\mu^*$ be the measure induced by this mapping:
 for a Borel set $E$, $\m^*(E)$ equals
 $\mu(\F^{-1}(E))$.
\begin{proposition}\label{4:Prop2} Suppose that the
 mapping $\F$ satisfies the condition
 \begin{equation}\label{4:Prop2Cond}
    |\F(w)-\l|\le a |w|^\t, \; \t<1
\end{equation}
for some $a\ge0$, for $|w|$ large enough. Let
$\mu^*_\pm$ also satisfy Condition \ref{7:ConditionC}.
Fix some $R>0$, such that \eqref{4:Prop2Cond} is
satisfied for $|w|>R$ and define the function
\begin{eqnarray}\label{4:Prop2:pot}
    \Psi^*(z)=\int\limits_{|w|<R}\Re\left[\ln(1-z/\F(w))-\ln(1-z/w)\right]d\m(w)+\nonumber\\
    \int\limits_{|w|\ge R}\Re\left[\ln(1-z/\F(w))-\ln(1-z/w)+z(\F(w)^{-1}
    -w^{-1})\right]d\m(w)
\end{eqnarray}
Then the function $\Psi^*$ satisfies the equation
$\Delta\Psi^*=2\pi(\mu^*-\mu)$ and, moreover
\begin{equation}\label{4:Prop2:potest}
    |\Psi^*(z)|\le C|z|^{1+\t}.
\end{equation}
\end{proposition}

\begin{proof} The fact that the integral converges is
obvious. To check that it satisfies the Poisson
equation, it is sufficient to notice  that for a
bounded domain $\Omega$ containing the point $z$,
\begin{equation}\label{4:changeOfVariiables}
    \int\limits_\Omega
    \Re\ln(1-z/\F(w))d\mu(w)=\int\limits_{\F(\Omega)}\Re\ln(1-z/w')d\mu^*(w').
\end{equation}

To prove the crucial inequality \eqref{4:Prop2:potest}
we have to estimate only the second integral in
\eqref{4:Prop2:pot} since the first one may grow at
most logarithmically at infinity.

For a fixed $z$, we split the plane into three regions
\begin{eqnarray*}\label{4:P2.1}
    \Omega_0=\{w\in \C,\; R<|w|\le|z|/2\},
\Omega_1=\{w\in\C,\; |z|/2<|w|\le 2|z|\},\\
\Omega_2=\{w\in\C,\; 2|z|<|w|\}.
\end{eqnarray*}
and estimate the corresponding integrals $I_0,
I_1,I_2$ separately.

For the integrand in $I_2$, we have
\begin{eqnarray}\label{4:P2.2}
    \!\!\!\!\!\!\!\!\!\!\!\ln\left(1-\frac{z}{\F(w)}\right)-\ln\left(1-\frac{z}{w}\right)
    +z\left(\frac1{\F(w)}-\frac1{w}\right)=\;\;\;\;\;\;\;\;\;\;\\
     \!\left[\left(\!\ln\!\left(\!1\!-\!\frac{z}{\F(w)}\!\right)\!+\!\frac {z}{\F(w)}\!+
     \!\frac
     {z^2}{2\F(w)^2}\!\right)\!
     -\!\left(\!\ln\left(\!1\!-\!\frac{z}{w}\!\right)\!+\!\frac{z}{w}\!+\!\frac
     {z^2}{2w^2}\right)\right]\!-\!\frac12\left[ \!\frac {z^2}{\F(w)^2}-\frac
     {z^2}{w^2}\!\right],\nonumber
\end{eqnarray}
and  correspondingly, $I_2$ splits into $I_{21}$ and
$I_{22}$. For $|w|>2|z|$, supposing $|z|$ is large
enough, we have $|z/\F(w)|<3/4$.

Consider the function $h(\xi)=\ln(1-\xi)+\xi+\xi^2/2$,
analytical in the open unit disk. For
$|\xi_1|,|\xi_2|\le\frac34$, we have
\begin{equation}
|h(\xi_1)-h(\xi_2)|\le|\xi_1-\xi_2|\max_{t\in[\xi_1,\xi_2]}|h'(t)|\le
4|\xi_1-\xi_2|
\max(|\xi_1|^2,|\xi_2|^2).\label{4:CorrL2.3}
\end{equation}
We set here $\xi_1=\frac{z}{\F(w)}$,
$\xi_1=\frac{z}{w}$, obtaining from \eqref{4:CorrL2.3}
\begin{eqnarray*}
    \left|h\left(\frac{z}{\F(w)}\right)-h\left(\frac{z}{w}\right)\right|\le 4
    \left|\frac{z}{\F(w)}-\frac{z}{w}\right|\max\left(\left|\frac{z}
    {\F(w)}\right|^2,\left|\frac{z}{w}\right|^2\right)\nonumber\\
    \le
    C|z|^3\left|\frac1{\F(w)}-\frac1{w}\right||w|^{-2}\le
    C|z|^3|w|^{-4+\t}.
\end{eqnarray*}
Therefore we get
\begin{equation}\label{4:CorrL2.5}
    |I_{21}|\le
    C|z|^3\int\limits_{|w|>2|z|}|w|^{-4+\t}d|\mu|(w)\le
    C |z|^{1+\t}.
\end{equation}

Further on, in order to estimate $I_{22}$, we make the
transformation $$\frac12\left(\frac{z}{w}\right)^{2}
-\frac12\left(\frac{z}{\F(w)}\right)^{2}=
\frac12z^2(w-\F(w))\frac{w+\F(w)}{w^2{\F(w)}^2},$$
which gives
$$\left|\frac12\left(\frac{z}{w}\right)^{2}-\frac12\left(\frac{z}{\F(w)}\right)^{2}
\right|\le C|z|^2|w|^{-3+\t},
$$
and therefore
\begin{equation}\label{4:CorrL2.6}
|I_{22}(z)|\le
C|z|^2\int_{\Omega_2}|w|^{-3+\t}d|\mu|(w)\le
Ca|z|^{1+\t}.
\end{equation}
Taken together, \eqref{4:CorrL2.5} and
\eqref{4:CorrL2.6} give
\begin{equation}\label{4:CorrL2.7}
    \left|I_{2}(z)\right|\le C |z|^{1+\t}.
\end{equation}

We pass to estimating $I_1$. We write the integrand
now as
\begin{eqnarray}\label{4:P2.4}
\Re\left[\ln(1-z/\F(w))-\ln(1-z/w)+z(\F(w)^{-1}-w^{-1})\right]=\nonumber\\
\ln|\F(w)/w|+\ln\left|\frac{\F(w)-z}{w-z}\right|+\Re(z(F(w)^{-1}-w^{-1})).
\end{eqnarray}
In the first term on the right in \eqref{4:P2.4}, we
have $|1-\F(w)/w|\le1/2$,  therefore
$|\ln|\F(w)/w||=|\ln|(1-(\F(w)-w)/w)||\le
C|w|^{-1+\t}$, and we get the estimate by $|z|^{1+\t}$
for the integral over $\Omega_1$. The integral over
$\Omega_1$ of the second term in \eqref{4:P2.4} we
split into the sum of $I_{11}$ and $I_{12}$, the
former being the integral over the domain
$\Omega_{11}: |w-z|\ge 4a|z|^\t$ and the latter over
$\Omega_{12}=\Omega_1\setminus\Omega_{11}.$ For
$I_{11}$, we have $\frac{|\F(w)-w|}{|w-z|}\le
\frac12$, so
$$I_{11}\le
C\int_{\Omega_{11}}\frac{|\F(w)-w|}{|w-z|}d\mu(w)\le
C|z|^\t\int_{\Omega_{11}}|w-z|^{-1} d\mu(w)\le C
|z|^{1+\t}.$$ For the integral over $\Omega_{12}$ we
apply the rough estimate
$|\ln\left|\frac{\F(w)-z}{w-z}\right||\le
|\ln\left|{\F(w)-z}\right||+|\ln\left|{w-z}\right||$
and integrate each term separately. Since the
integration here is performed over the disk with
radius $4a|z|^\t$, the conditions on the measures
$\mu$ and $\mu^*$ imply that $I_{12}$ can be estimated
by $|z|^{2\t}|\ln|z||=O(|z|^{1+\t})$.

  Finally, for
$I_0$  note that for $w\in\Omega_0$ we have
$\left|\frac{\F(w)-w}{w-z}\right|\le
\frac{2a|w|^\t}{|z|}\le 1/2$ for $|z|$ sufficiently
large, therefore
\begin{eqnarray}
 \left|I_0(z)\right|\le\int_{\Omega_0}|\ln|\F(w)w^{-1}||d|\mu|(w)
 +\int_{\Omega_0}|\ln|\frac{\F(w)-z}{w-z}||d|\mu|(w)\nonumber\\+|z|\int_{\Omega_0}
 |w^{-1}-\F(w)^{-1}|d|\mu|(w)
 \le
 C\int_{\Omega_0}|w|^{-1+\t}d|\mu|(w)\nonumber\\+C\int_{\Omega_0}|w|^\t|w-z|^{-1}d|\mu|(w)+
 C|z|\int_{\Omega_0}|w|^{-1+\t}d|\mu|(w).
\end{eqnarray}
All integrals in this expression are estimated by
$CA|z|^{1+\t}$.
\end{proof}

Following the general pattern described in the
beginning of the Section, we arrive at the following
case where the infiniteness of zero modes is granted.
Note that the somewhat complicated conditions imposed
on the unperturbed measure are aimed to cover the
interesting cases described in the beginning of the
Section.

\begin{theorem}\label{4:TheoremContinRearr} Let
$\mu_0$ be a measure, with discrete part
$\mu_{0,disc}=\sum_{\l\in\L_0}\a_\l \d(z-\l)$
supported on the set $\L_0$ satisfying
\eqref{2:sparse}, with $0<\a_\l\le\theta_0<1$ for the
operator $\Pc_{\max}$ and $\a_\l\in[-1/2,1/2)$ for
$\Pc_{EV}$. Let $\l_0(z)$ be the point in $\L_0$
closest to $z$ (or any of such points) and $\a_0(z)$
be $\a_{\l_0(z)}$. Suppose that the measure $\mu_0$
 admits
a potential $\Psi_0$ subject to the estimate
\begin{equation}\label{4:unpert}
    \Psi_0(z)-\a(z)\ln|z-\l_0(z)|\ge \g|z|^2, \g>0
\end{equation}
 for sufficiently large $|z|$ (if $\L$
is empty, $\ln|z-\l(z)|$ in \eqref{4:unpert} is
replaced by zero). Let $\mu $ be another measure and
$\F$ be a Borel measurable mapping so that the
conditions of Proposition~\ref{4:Prop2} are satisfied.
Then for the measure $\mu_0+\mu-\mu*$ there exists a
potential  $\Psi^*$ also satisfying \eqref{4:unpert}
with some $\g', 0<\g'<\g$, and thus there are
infinitely many zero modes for the Pauli operator
$\Pc_{-\max}$, resp., $\Pc_{-EV}$ with the field
$\mu_0+\mu-\mu*$.\end{theorem}

The version of Proposition~\ref{4:Prop2}  for the case
of a discrete perturbing measure $\mu$ is also valid.
We only give the formulation here, the proof being
practically  the same as above.

\begin{proposition}\label{4:Proprearrdisc}
Let $\mu$ be a discrete measure
$\mu=2\pi\sum_{\l\in\L}\a_\l\d(z-\l)$ such that
   $|\a_\l|\le\theta_0<1$
and the discrete set $\L$, the support of $\m$,
 satisfies the uniform discreteness condition \eqref{2:sparse}. Let
 each point $\l\in\L$ move to a new position $\l'$ so
 that
the mapping $\F:\l\mapsto\l'$ transforms $\L$ to
another discrete set $\L'$, also satisfying
\eqref{2:sparse}, moreover, for any $\l'\in\L'$,
$$
  |\a'(\l')|\equiv \left|\sum_{\l\in\F^{-1}(\l')}\a_\l\right|\le\theta_0.
$$
Suppose finally that $\F$ satisfies
\eqref{4:Prop2Cond}. We define the measure
$\mu'=2\pi\sum_{\l'\in\L'}\a'_\l\d(z-\l')$ and the
function
\begin{equation}\label{4:discpoten}
    \Psi(z)=\frac{1}{2\pi}\sum_{\l'\in\L'}\sum_{\l\in\F^{-1}(\l')}\a_\l
    \left[\ln|\l'-z)|-\ln|\l-z|+
    \Re(z(\l'^{-1}-\l^{-1}))\right].
\end{equation}
Define also $\a(z)$ as $\a_{\l(z)}$ where $\l(z)$ is
the point in $\L$ , closest to $z$, and, similarly
$\a'(z)=\a'(\l'(z))$ where $\l'(z)$ is the point in
$\L'$ , closest to $z$. Then the function $\Psi$
satisfies the Poisson equation $\Delta\Psi=\mu'-\mu$
and
\begin{equation}\label{4:EstimDisc}
    \left|\Psi(z)-\a'(z)\ln|z-\l'(z)|+\a(z)\ln|z-\l(z)|\right|\le
    C|z|^{1+\t}.
\end{equation}
\end{proposition}

The discrete version of
Theorem~\ref{4:TheoremContinRearr} is now formulated
as following.
\begin{theorem}\label{4:Theorem}
Suppose that $\mu_0$ is a measure satisfying
conditions of Theorem~\ref{4:TheoremContinRearr}. Let
$\mu$ be a discrete measure and $\F$ a mapping such
that the conditions of
Proposition~\ref{4:Proprearrdisc} are satisfied.
Suppose finally that the union of the sets
$\L_0,\L,\L'$ also satisfies \eqref{2:sparse}.

Then the measure $\mu_0+\mu'-\mu$ possesses the
potential $U(z)$ subject to
\begin{equation}\label{4:bigdiscrete}
    U(z)-\a_0(z)\ln|z-\l_0(z)|+\a'(z)\ln|z-\l'(z)|-\a(z)\ln|z-\l(z)|\ge\g'|z|^2,\;
    0<\g'<\g
\end{equation}
for $|z|$ large enough, and thus the Pauli operators
$\Pc_{-\max}$, resp., $\Pc_{-\max}$ have infinitely
many zero modes.\end{theorem}
We will give later some
examples how Theorems \ref{4:TheoremContinRearr} and
\ref{4:Theorem} can be applied in interesting concrete
situations.

\bigskip

Further on  we show that the quadratic growth of the
potential and the property of the null subspace to be
infinite-dimensional are stable also under certain
additive perturbations of the magnetic field. The
perturbation of the field is a signed  measure $\m$
such that its discrete part $\m_{\disc}$ with weights
$\b_{\l}$ satisfies conditions of
Proposition~\ref{4:Proprearrdisc}
 and the continuous part $\mu_{\cont}$ satisfies
 \begin{equation}
\label{5:uniformEstimate}\int\limits_{|w-z|<r^0}|\ln|w-z||d|\mu_c|(w)\le
B
\end{equation}
for some  $B$, for all $z\in\C$ ($r^0$ is the constant
in \eqref{2:sparse}). To evaluate the size of the
measure $\mu$, the following characteristics will be
used:
$$\omega(r)=\omega(r,\mu)=|\mu|(\Db(0,r)),$$
where, recall, $\Db(0,r)$ is the disk with center at
the origin and radius $r$, and
  $$  M(r)=\int_{r_1\le|w|\le r}w^{-2}d\mu(w)$$
for some fixed $r_1$. To such measure $\mu$  we
associate
 the potential $\Psi_\mu(z)$ defined as
\begin{equation}\label{5:Perturbed Potential}
    \frac{1}{2\pi}\left[\int\limits_{|w|\le
    R}\!\!\ln|1-z/w|d\nu(w)\!+\!\int\limits_{|w|>R}\!\!\left(\ln\left|1-\frac{z}{w}\right|+
    \Re\left(\frac{z}{w}+\frac12\left(\frac{z}{w}\right)^2\right)\right)d\nu(w)\right].
\end{equation}
The value of $R$ will be chosen later. Note, moreover,
that changing $R$, we add  a harmonic function, the
real part of a second degree polynomial to $\Psi_\mu$.
For a given point $z\in\C$, there may be only one
point $\l\in\L$ in the $r^0/2$-neighborhood of $z$. If
such point exists we denote it by $\l(z)$ and set
$\b(z)=\b_{\l(z)}$ and $L(z)=\b(z)\ln|z-\l(z)|$,
otherwise $L(z)$ is set to be zero.

Now we formulate our main estimate.
\begin{proposition}\label{5:PropMainEstimate}
Under the above conditions the following estimates
hold.
\begin{enumerate}
\item Suppose that for $r$ large enough
\begin{equation}\label{5:PropEstforOmega}
\omega(r)\le C(r^{2-\t}), \t>0.
\end{equation}
Then 
 $$   |\Psi_\mu(z)-L(z)|\le C'(|z|+1)^{2-\t}\ln|z|.$$

\item
 Suppose that for $r$ large enough,
\begin{equation}\label{5:PropEstforOmega2}
\omega(r)\le \epsilon r^{2}
\end{equation}
 and
\begin{equation}\label{5:EstimateForM}
    |M(r)|\le\epsilon , r>R_0.
\end{equation}
Then
 \begin{equation}\label{5:Propresult forM}
   |\Psi_\mu(z)-L(z)|\le C\e\ln(1/\e)(|z|+1)^2.
\end{equation}
\end{enumerate}
\end{proposition}
Note  that in both cases the estimate can be written
as
$$|\Psi_\mu(z)-L(z)|\le
C\tilde{\omega}(|z|)\ln(\sqrt{\tilde{\omega}(|z|)}|z|^{-1}),$$
where $\tilde{\omega}(r) $ is the majorant for
$\omega(r)$ in the proposition.

Before starting to prove the Proposition, we explain,
that, in the first case,  the measure $\mu$ is
supposed to be rather sparse, so that  the
$|\mu|$--measure of the disk $\Db(0,r)$ grows slower
than the area of the disk. In the second case, the
measure of the disk may grow as fast as its area, with
a small factor, however an additional condition
\eqref{5:EstimateForM} is imposed. This latter
condition is satisfied if the measure $\mu$ is more or
less uniformly distributed in all directions or at
least, has zero second circular harmonic.  On the
other  hand, if $\mu$ is the Lebesgue measure
restricted to some angle in $\C$ not coinciding with
the whole plane or a half-plane, then the condition
\eqref{5:EstimateForM} is violated.

\begin{proof} We prove the parts (1) and (2) of the proposition
simultaneously. Similar to the proof of
Proposition~\ref{4:Proprearrdisc}, we divide the plane
with the disk $|w|<R$ removed into the same three
parts, $\Omega_0=\{R<|w|<\frac12|z|\}$,
$\Omega_1=\{\frac12|z|\le|w|<2|z|\}$, and
$\Omega_2=\{|w|\ge\frac12|z|\}$. Correspondingly, the
integral in \eqref{5:Perturbed Potential} splits into
the sum of three integral which we denote by
$I_0,I_1,I_2$.

For $I_0$ we have
\begin{eqnarray}\label{5:I0:1}
|I_0|\!=\!\left|\int\limits_{\Omega_0}\ln|1-z/w|d\mu(w)+\!\!\int\limits_{\Omega_0}
\Re(z/w)d\mu(w)
+\frac12\int\limits_{\Omega_0}\Re\left((z/w)^2\right)d\mu(w)\right|\nonumber \\
\le
\int\limits_{\Omega_0}\!|\ln|1-z/w||d|\mu|(w)+\left(\int\limits_{\Omega_0}\!|z/w|d|\mu|(w)+
\frac12\left|\Re\left(\!z^2\!\int\limits_{\Omega_0}w^{-2}d\mu(w)\right)\right|\right).
\end{eqnarray}

For $w\in\Omega_0$, we have $|z-w|\ge|z|/2$ and
$|1\le|1-z/w|\le|2z/w|$, and therefore
\begin{eqnarray}\label{5:I0:2}
\int\limits_{\Omega_0}|\ln|1-\frac{z}{w}||d\mu(w)\le
\int\limits_{\Omega_0}\ln|\frac{2z}{w}|d\mu(w)
=\int\limits_R^{|z|/2}\ln|\frac{2z}{w}|d\omega(r)\nonumber\\=2\pi
(\ln4\;\omega(\frac{|z|}2)-\ln(\frac{2|z|}{R})\omega(r))+2\pi\int\limits_R^{|z|/2}\frac{r}{2|z|}\omega(r)dr.
\end{eqnarray}
The conditions \eqref{5:PropEstforOmega}, resp.,
\eqref{5:PropEstforOmega2}, imply then that the
expression in \eqref{5:I0:2} is majorated by
$C\tilde\omega(|z|))$.

 The second term on the right-hand side in
\eqref{5:I0:1} is estimated simply by
$$\int\limits_{|w|<\frac12|z|}\Re(\frac{z}{w})d\mu(w)\le
C|z|\int\limits_R^{|z|/2}r^{-1}d\omega(r)\le
C\tilde\omega(|z|).$$ It is in the last term that the
treatment of
 two cases is different. In the case (1),
    $$\left|\Re\left(z^2\int\limits_{\Omega_0}w^{-2}d\mu(w)\right)\right|\le
    C|z|^2 \int_R^{|z|/2} |r|^{-2}d\omega(r)\le C
    r^{2-\t}.$$
In the case (2) the estimate by absolute value is not
sufficient, so we act differently:
$$\left|\Re\left(z^2\int\limits_{\Omega_0}w^{-2}d\mu(w)\right)\right|\le|z|^2
\left|\int\limits_{\Omega_0}w^{-2}d\mu(w)\right|\le
|z|^2|M(|z|/2)|\le \e|z|^2.$$
Taken together, the last three inequalities give
   $ |I_0|\le C\tilde{\omega}(|z|).$

To estimate $I_1$, we split the integration set
$\Omega_1$ into the disk $\Db=\{|w-z|<r^0/4\}$ and
$\Omega_1'=\Omega_1\setminus\Db$, correspondingly,
$I_1=I_{10}+I_1'$. If there are no points of $\L$ in
the $r^0/2$ -neighborhood of $z$, so that the measure
$\mu$ is continuous in $\Db$, then
 $I_{10}$ is estimated by a constant by
\eqref{5:uniformEstimate}. If there is a (unique)
point  $\l(z)\in\L$ in the above disk, then this
contribution equals $\b(z)\ln|z-\l(z)|$ plus some
bounded term coming from the continuous part of the
measure. In the notation of the Proposition, in any
case,
$ |I_{10}  -L(z)|\le
    C.$

In the set $\Omega_1'$, outside the disk $\Db$, we
again estimate the absolute value of the integral by
\begin{equation}\label{5:I1:2}
|I_1'|\le \int_{\Omega_1'}
|\ln|1-z/w|d\|\mu|(w)+\int_{\Omega_1'}|z/w|d|\mu|(w)
+\left|\frac12\int_{\Omega_1'}z^2/w^2d\mu(w)\right|.
\end{equation}
The second and the third terms in \eqref{5:I1:2} are
estimated by $|z|^{2-\t}$ in the same way as the
similar terms in $I_0$.

More trouble we have with the first term in
\eqref{5:I1:2}, and the estimates we get for this term
are the worst ones. We have $|\ln|1-z/w||\le
C+|\ln\left|\frac{z-w}{z}\right||$ and the first term
here contributes  with $O(\omega(|z|))$ to $I_1'$. We
also have
$$\int\limits_{z\in\Omega_1',|z-w|\ge|z|/2}\left|\ln\left|\frac{z-w}{z}\right|\right|\le
C \omega(|z|).$$ So it remains to evaluate
$$I_{11}=\int\limits_{r^0/4<|w-z|\le|z|/2}|\ln|(z-w)/z||d|\mu|(w).$$
We split the region $r^0/4<|w-z|\le|z|/2$ into annuli
$U_k=\{kr^0/4<|z-w|\le(k+1)r^0/4\}$, k=1,2,... (the
last one may be somewhat larger). Denote by $S_k$ the
$|\mu|$-measure of the annulus $U_k$. We have, of
course,
 $$   \sum S_k\le |\mu|(\Omega_1')\le \omega(2|z|).$$
At the same time, since each annulus $U_k$ can be
covered by no more than $8k\pi$ disks with radius
$r^0/4$, the condition \eqref{5:uniformEstimate}
implies that $S_k\le \g k$, $\g=8\pi B_0$. Now, we
majorize the integral $I_{11}$ by the sum of integrals
over annuli $U_k$ and estimate it from above in the
terms of $S_k$:
\begin{equation}\label{5:I1:4a}
I_{11}\le\sum \ln\left(\frac{4|z|}{kr_0}\right)S_k.
\end{equation}
The sequence $\ln\left(\frac{4|z|}{kr_0}\right)$
decreases in $k$. Therefore, if we replace $S_1$ by
its largest possible value $\g$ with simultaneous
decreasing of the rest of $S_k$ to some new
non-negative values, keeping the sum the same, the sum
in \eqref{5:I1:4a} can only increase. We perform the
same operation with $S_2$, making it equal $2\g$, then
with $S_3$ and so on, until all 'small' $S_k$, $k<N$
have their maximal possible values, $S_k=k\g $, the
next one, $S_N$ is smaller than $N\g$, and the rest
are zeros. Since the sum of $S_k$ is not greater than
$\omega(2|z|)$, we have $\g N(N+1)/2\le \omega(2|z|)$,
so $N\le (2\omega(2|z|)/\g)^{1/2}$. In this way we
have reduced the task of  estimating $I_{11}$ to
evaluating the sum
$$\g\sum_{k=1}^{ (2\omega(2|z|)/\g)^{1/2}}
k\ln\left(\frac{4|z|}{kr^0}\right).$$ This sum can be
estimated by the integral
$$\g\int_{r^0/4}^{
(2\omega(2|z|)/\g)^{1/2}}t\ln\left(\frac{4|z|}{tr_0}\right)dt.$$
After the substitution $t=s4|z|/r^0$, the integral
transforms to
$$C\g|z|^2\int_0^{(2\omega(2|z|)/\g)^{1/2}/|z|}s\ln(1/s)ds,$$ which is
estimated directly since $\int_0^\e s\ln(1/s)ds
=O(\e^2\ln(1/\e)), \e\to 0.$ Thus we obtain
$$    I_{11}\le C
    \omega(|z|)\ln(\sqrt{\omega(|z|)}/|z|).$$
Collecting this inequality with previously found
estimates for other contributions to $I_1$, we obtain
  $$  |I_1(z)-L(z)|\le C
    \omega(|z|)\ln(\sqrt{\omega(|z|)}/|z|).$$
The term $I_2$ is the easiest one. Since $|z/w|<1/2$
in $\Omega_2$, we have
$$\left| \ln|1-z/w|+\Re(z/w+(z/w)^2/2)\right|\le \frac23 \left|{z}/{w}\right|^3.$$
Therefore
   $$ |I_2|\le \frac23
    |z|^3\int_{\Omega_2}|w|^{-3}d|\mu|(w)\le
    C|z|^3\int_{2|z|}^\infty
    r^{-4}\tilde{\omega}(r)dr,$$
which gives the required estimate in both cases.
\end{proof}
The Proposition we  have proved leads  to the
following Theorem.
\begin{theorem}\label{5:MainTheorem} Let $\mu_0$ be a
 measure subject to conditions of
 Theorem~\ref{4:TheoremContinRearr}.
Suppose that the perturbation $\mu$ satisfies
conditions of Proposition~\ref{5:PropMainEstimate}. We
suppose also that the number $\e$ in the case (2) of
the Theorem is small enough, so that
$C_\mu\e\ln(1/\e)<\g$, where $C_\nu$ is the constant
in \eqref{5:Propresult forM}.  Suppose finally that
the union of the discrete sets $\L_0$ and $\L$ satisfy
the condition \eqref{2:sparse}, probably, with
different $r^0$. Then the measure $\mu^*=\mu_0+\mu$
possesses a potential $\Psi^*$ satisfying
\begin{equation}\label{5:MainTheoremEstimate}
    \Psi^*(z)-\alpha^*(z)\ln(d^*(z))\ge \g^*|z|^2, \;
    \g^*>0
\end{equation}
where $d^*(z)$ is the distance from $z$ to the nearest
point of the support of the discrete part of measure
$\mu^*$ and $\a^*(z)$ is the measure $\mu$ of this
point (or any of such points, if there are several of
them). In this situation the Pauli operators
$\Pc_{-\max}$ resp. $\Pc_{-EV}$ have infinitely many
zero modes\end{theorem}

Although Theorem~\ref{7:MainTheorem} provides us with
very general conditions for infiniteness of zero
modes, it applies only in such cases when there are no
arbitrarily large regions in the plane  where the
field is negative - this would destroy the
subharmonicity of the averaged potential. The
perturbation results enable us to establish the
infiniteness of zero modes in certain situations when
such regions are present: they arise as a result of
perturbations allowed by the theorems in this section.

The starting point has to be a magnetic field where an
exponential estimate of the form
\eqref{4:TheoremContinRearr} is already known. After
this, we can apply allowed types of perturbation again
and again, as long as at each step the perturbation
satisfies the general conditions of this Section, in
other words, is weak enough, in the proper sense.

The examples below do not exhaust all kinds of
perturbations given by our theorems but rather
illustrate their possibilities.

\begin{example}\label{6:Ex1} Let $\mu_0$ be
a measure for which the conditions of
Theorem~\ref{5:MainTheorem} are satisfied. Let
$\Omega$ be a set in $\C$ such that the
$|\mu_0|(\Omega\cap\Db(0,R))=O(R^{2-\t})$ for some
$\t>0$ and $\mu_\Omega$ be the measure $\mu_0 $
restricted to $\Omega$. Then for the magnetic field
$\mu_0-B\mu_\Omega$, for any positive $S$ the operator
Pauli has infinitely many zero modes. In fact,
Theorem~\ref{5:MainTheorem}, part (1) applies here. If
$B>1$, the measure $\mu_0-B\mu_\Omega$ has in $\Omega$
the sign opposite to the one of $\mu_0$. The domain
$\Omega$ can be rather large. In particular, for the
examples of $\mu_0$ discussed above, the set $\Omega$
may be the domain $\{|x_2|\le C(1+|x_1|)^\t\}$, with
some $\t<1$, thus admitting  rather large
negative-field regions. Another case is $\Omega$ being
the union of disks $\Db_k$ of radii $R_k$ tending to
infinity as $k\to\infty$, such that the combined area
of the disks $\Db_k$ fitting into the disk $\Db(0,R)$
is majorated by $R^{2-\t}$.\end{example}

If the field to be perturbed is more regular, even
stronger perturbations are allowed.

\begin{example}\label{6:Ex2} Let $\mu_0$ be a regular lattice
of Aharonov-Bohm potentials, possibly on the
background of the constant magnetic field, or a
periodic (quasi-periodic) measure satisfying
Condition~\ref{7:ConditionC} with nonzero $\sum_k
\frac{\Phi_k}{|\digamma_k|}$, so that the potential
with quadratic growth exists.
 Let $\Omega=\Omega_-\cup\Omega_+$ be the
double angle in the plane,
$$\Omega=\{\arg z\in(\theta_1,\theta_2)\}\cup \{\arg
z\in(\theta_1+\pi,\theta_2+\pi)\},$$ and
$\mu_{\Omega_\pm}$ be the restriction of $\mu_0$ to
the angles $\Omega_\pm$. Then, in the notations of
Example~\ref{6:Ex1}, Theorem~\ref{5:MainTheorem}, part
(2), establishes the required estimate for the
potential of the measure
$\mu-B\mu_{\Omega_+}+B\mu_{\Omega_-}$ and therefore
the infiniteness of zero modes for the Pauli operator,
as soon as the size of the angle, $\theta_2-\theta_1$,
is small enough. This smallness guarantees the
fulfillment of the condition
\eqref{5:PropEstforOmega2}, while the symmetry of the
domain $\Omega$ leads to fulfillment of
\eqref{5:EstimateForM}. Instead of symmetry, we can
require that the domain $\Omega$ consists of four
angles and is invariant with respect to rotation by
$\pi/2$. Then, if the size of the angles is small
enough, the perturbation $-B\mu_\Omega$ again
satisfies conditions of the perturbation theorems.  Of
course, one can change the field not necessarily in
two (four) angles but in any domain symmetric with
respect to the rotation by $\pi$, as long as the area
of the portion of the domain in the disk $\Db(0,R)$
grows not faster than $\e R^2$, with $\e$ small
enough.
\end{example}

The most efficient pattern for applying the
re-arrangement Theorems~\ref{4:TheoremContinRearr} and
\ref{4:Theorem} is the following. Suppose that we have
a decomposition of the plane $\C$ into the union of
disjoint sets $\Omega_j$ so that the diameter
   of $\Omega_j$ is not greater than $C|z_j|^{\t},\;\t<1$,
   where $z_j$ is some point in $\Omega_j$. Let
   $\mu$ a perturbing measure satisfying conditions of
    Theorems~\ref{4:TheoremContinRearr} or
\ref{4:Theorem},  such that $\mu(\Omega_j)=0$. Then we
can define the mapping $\F$ as $\F(\Omega_j)=\{z_j\}$,
so that the whole set $\Omega_j$ is mapped into one
point. This mapping induces the zero measure $\mu^*$.
The perturbation theorems immediately produce the
estimates for the potential of the measure $\mu$
together with infiniteness of zero modes. This
reasoning illustrates that positive and negative parts
of the perturbation can cancel each other even if they
lie rather far apart.

The sets $\Omega_j$ can be chosen rather arbitrarily.
The following construction may be useful.

\begin{lemma}\label{6:Lemma} Let $\t\in(0,1)$. Then
there exists a family of squares $\Qb_j$, with centers
at $z_j$ and sides $d_j$ covering the plane and having
no common interior points, such that
\begin{equation}\label{6:Lemma.1}
    c|z_j|^\t\le d_j\le C|z_j|^\t,
\end{equation}
with some constants $c,C$, for all squares, except the
one containing the origin.\end{lemma}

\begin{proof} We start with the unit square $\Qb_0$
centered at the origin. surround $\Qb_0$ by eight
equal squares which will be called the first layer, so
the first nine squares form a square with side $3$,
and condition~\eqref{6:Lemma.1} is satisfied, with,
say, constants $c=1/4, C=4$. Further, inductively,
having already a square in the plane filled with
squares $\Qb_j$, so that the inequalities
\eqref{6:Lemma.1}, we surround this square by squares
of the same size as in the last layer, if for the new
layer \eqref{6:Lemma.1} still holds. In the opposite
case, if there are  $N$ squares in the last layer, the
new layer will be composed by the squares with the
side twice as large as in the last layer (if $N$ is
even) or with the side $2N/(N+1)$ times larger (if $N$
is odd). One can easily check that this construction
preserves the inequality \eqref{6:Lemma.1}, and
repeating it we get the required covering.\end{proof}

To show how the above construction works, we consider
the case of a continuous measure.

\begin{example}\label{6:Ex4} Let $\mu$ be a continuous
measure satisfying the conditions of
Theorem~\ref{4:TheoremContinRearr} with the following
property:  for any square $\Qb$ with center at $z$, as
in Lemma~\ref{6:Lemma}, $\mu(\Qb)-B|\Qb|\le
C|z|^{2\t-\e}$, for some $\e>0$ with a positive
constant $B$, $|\Qb|$ denoting the area of the square.
We set $\mu_0$ being $B$ times the Lebesgue measure;
it possesses the required quadratically growing
potential and will serve as the unperturbed field. The
difference, $\mu-\mu_0$ will be represented as a sum
of two terms, $\mu_1+\mu_2$. We set
$\mu_1=\mu-\mu(\Qb_j)/|\Qb_j|dx$ on the square
$\Qb_j$, and $\mu_2=(\mu(\Qb_j)/|\Qb_j|-B)dx$. Then
the perturbation $\mu_1$ satisfies conditions of
Theorem~\ref{4:TheoremContinRearr}, with $\F$ mapping
the whole square $\Qb_j$ to the single point
$z_j\in\Qb_j$, and $\mu_2$ satisfies conditions of the
Theorem~\ref{5:MainTheorem}.

Note that in the situation of the example, the regions
where the field points in the 'wrong direction', can
be very large. In fact, one can choose $\mu$ so that
for each square  $\Qb_j$ in our covering, $\mu>0$ in
the narrow strip near the boundary, with width
$a|z_j|^\t$, with an arbitrarily small fixed $a$, and
$\mu$ is negative in the main part of the
square.\end{example}

A similar property holds for discrete measures
\begin{example} Let $\L$ be a regular lattice and
$\a_\l,\;\l\in\L$ is the collection of intensities,
$\a_\l\in[-1/2,1/2)$. We suppose that for any square
$\Qb$ with center at $z$ (or only for squares $\Qb_j$
constructed in Lemma~\ref{6:Lemma}),
$$\left|\sum_{\l\in\L\cap\Qb}\a_\l-B|\Qb|\right|\le C|z|^{2\t-\e}, B>0$$
for some $\e>0$. Then, similarly to the previous
example, the system of $AB$ solenoids with intensities
$\a_\l$ placed at the points of $\L$ can be obtained
by a re-arrangement perturbation and an additive
perturbation of the regular AB lattice with equal
intensities. This proves infiniteness of zero modes
for the operator $\Pc_{-EV}$ for such configuration of
the field. Moreover, we can even suppose that
initially $\L$ is not a regular lattice but just a
discrete set, in proper sense, almost uniformly
distributed in the plane. This situation is taken care
of by additional perturbation consisting in moving the
points of $\L$ to the points of a regular lattice and
then making one more additive perturbation to dispose
of the points which cannot be moved. We do not go into
details here.\end{example}

 \end{document}